\documentclass{article}

\usepackage{PRIMEarxiv}

\usepackage{lineno,hyperref}
\usepackage{amsmath}
\usepackage{amsfonts}
\usepackage{amssymb}
\usepackage{txfonts}
\usepackage{graphics}
\usepackage{epsfig}
\usepackage{bm}

\usepackage{algorithm}
\PassOptionsToPackage{noend}{algpseudocode}
\usepackage{algpseudocode}

\errorcontextlines\maxdimen

\usepackage[utf8]{inputenc} % allow utf-8 input
\usepackage[T1]{fontenc}    % use 8-bit T1 fonts
\usepackage{hyperref}       % hyperlinks
\usepackage{url}            % simple URL typesetting
\usepackage{booktabs}       % professional-quality tables
\usepackage{multirow}
\usepackage{nicefrac}       % compact symbols for 1/2, etc.
\usepackage{microtype}      % microtypography
\usepackage{fancyhdr}       % header

\title{Learning governing physics from output only measurements}

\author{Tapas Tripura\\
  Department of Applied Mechanics\\
  Indian Institute of Technology Delhi\\
  \texttt{tapas.t@am.iitd.ac.in} \\
  \And
      Souvik Chakraborty \\
  Department of Applied Mechanics\\
  School of Artificial Intelligence (ScAI)\\
  Indian Institute of Technology Delhi\\
  \texttt{souvik@am.iitd.ac.in} \\
}

\begin{document}
\maketitle

\begin{abstract}
Extracting governing physics from data is a key challenge in many areas of science and technology. The existing techniques for equations discovery are dependent on both input and state measurements; however, in practice, we only have access to the output measurements only. We here propose a novel framework for learning governing physics of dynamical system from output only measurements; this essentially transfers the physics discovery problem from the deterministic to the stochastic domain. The proposed approach models the input as a stochastic process and blends concepts of stochastic calculus, sparse learning algorithms, and Bayesian statistics. In particular, we combine sparsity promoting spike and slab prior, Bayes law, and Euler Maruyama scheme to identify the governing physics from data. The resulting model is highly efficient and works with sparse, noisy, and incomplete output measurements. The efficacy and robustness of the proposed approach is illustrated on several numerical examples involving both complete and partial state measurements. The results obtained indicate the potential of the proposed approach in identifying governing physics from output only measurement.
\end{abstract}

% keywords can be removed

\section{Introduction}
Physical systems are governed by the physical laws, often represented either in form of ordinary differential equations (ODE) or partial differential equations (PDE). The literature on techniques for the solution of ODE/PDE is quite matured and given sufficient computational resources, it is possible to solve the governing ODE/PDE with sufficient accuracy. Modern techniques such as physics-informed deep learning \cite{goswami2020transfer,raissi2018deep,chakraborty2021transfer,chakraborty2020simulation} and neural network based operator learning \cite{lu2019deeponet,li2020fourier} can also be used. However, due to various assumptions and approximations, the governing equations often fails to represent the exact physics of the system and results in modelling and prediction errors. With advancement in the sensor technology and the internet of things (IoT), we have access to data in abundance whereas the underlying governing physics often remains elusive specifically is domains such as climate science \cite{zanna2020data}, biology \cite{zheng2020maximum}, physics \cite{jia2017emergent}, chemistry \cite{noe2017collective} and finance \cite{zhang2006information} to name a few. One possible alternative is obviously to train a conventional data-driven machine learning model to obtain an input-output mapping; however, such models often do not generalize to unseen environment and out-of-distribution inputs. To address this issue, methods for data-driven equation discovery has recently been proposed. A seminal work in this area was carried out by \cite{bongard2007automated} wherein symbolic regression was utilized for discovering underlying structure of the data. Another important work in this area includes work by Brunton \textit{et al.}\cite{brunton2016discovering}. However, most of the work in this area is limited by the fact that information on both input and state vector are needed. In practice, the input is often not measurable and only few states are accurately observable; this greatly limits the applicability of the available physics discovery techniques. To address this issue, we propose a novel approach that leverages sparse learning, Bayesian statistics, and stochastic calculus and enables discovery of governing physics from output only measurements.

The development in the equation discovery techniques has made significant progress from early 1970s equation discovery was mostly dependent on the expertise in the interested area, to the modern machine learning techniques that can extract the physical law of the underlying process with only little human supervision. In the early 1970s the models were selected by maintaining a balance between the model complexity and its fitness. The fitness of the models were measured using Akaike (AIC) and Bayesian information criterion (BIC) \cite{akaike1974new,schwarz1978estimating}. Later advances in the machine learning tools in combination with new sophisticated data measurement instruments gave rise to the data-driven techniques for discovery of governing physics. Towards this, first attempt was made by combining symbolic regression with genetic programming to select the best combination of the functions from candidate functional forms that accurately represent the data \cite{bongard2007automated,schmidt2009distilling}. There are also equation-free methods that bypasses the need of formulating constitutive equations to track the time evolution of the dynamical systems in closed form. These methods provide a practical recipe for multiscale simulations through local microscopic simulations in time and space \cite{kevrekidis2009equation}. 

Following the work of symbolic regression \cite{bongard2007automated,schmidt2009distilling}, Sparse Identification of Nonlinear Dynamics (SINDy) was proposed for discovery of the governing physics of non-linear dynamical systems \cite{brunton2016discovering}. In terms of accuracy and intepretability of the discovered physics, the SINDy framework overcame the shortcomings of the previous discovery techniques. The idea behind SINDy was to use sparse linear regression in the purview of least-square fitting for selecting the most dominating candidate functions that best represent the data. Due to sparsity promoting approach, SINDy proved to be computationally efficient and scalable with the increase in the input dimension. In the later years, SINDy has shown tremendous applications and area-specific developments of sparse linear regression in various fields. Few examples are, sparse identification of biological networks in biology \cite{mangan2016inferring} and sparse identification of chemical reaction in chemistry \cite{hoffmann2019reactive,bhadriraju2019machine}, sparse modelling for state estimation in fluid mechanics \cite{loiseau2018constrained,loiseau2018sparse}, system identification of structures with hysteresis and sparse learning of aerodynamics of bridges in structural systems \cite{lai2019sparse,li2019discovering}, sparse model selection using an integral formulation \cite{schaeffer2017sparse}, sparse model selection of dynamical system using information criteria \cite{mangan2017model}, sparse identification for predictive control \cite{kaiser2018sparse}, identification of structured dynamical systems with limited data \cite{schaeffer2020extracting}, model identification using recovery of differential equations from short impulse response time-series data \cite{stender2019recovery}, identifying stochastic dynamic equation \cite{boninsegna2018sparse}, and discovery of partial differential equations \cite{rudy2017data,zhang2018robust}, among others.

Apart from SINDy, identification of non-linear dynamical systems using deep neural networks were also proposed by the researchers \cite{rudy2019deep,raissi2018deep,raissi2018multistep}. The deep learning approaches for non-linear dynamical systems exists as a black-box which takes the data as a input and provides the prediction as output. Later, a different approach within the Bayesian framework for discovery of governing physics from noisy and/or limited data was proposed \cite{fuentes2021equation,nayek2021spike,gupta2021bayesian}. In this approach the least-square based sparse linear regression was replaced by more accurate sparse Bayesian linear regression technique. Due to the ability of the Bayesian inference to perform simultaneous model selection and parameter estimation, this approach eliminated the possibility of overfitting of the governing model. The sparsity in the solution was promoted by assigning the appropriate sparse promoting priors over the weight vector of the sparse regression problem. This approach provided a natural elimination of the basis functions that do not present the data accurately. As a output, this framework provides the posterior distributions of the weight vector; this means one could accurately identify the uncertainty in the parameters and construct a confidence interval for future predictions.

Despite the progress in equation discovery from data in recent times, almost all the approaches require measurements for both state and input variables; the only exceptions is perhaps the Stochastic SINDy \cite{boninsegna2018sparse}. Unfortunately, we often only have access to the state variables only; this significantly limits the applicability of the available approaches. We hereby propose an equation discovery framework rooted in stochastic calculus, sparse learning, and Bayesian statistics, that require only the state estimates. The basic premise here is to treat the unknown input as a stochastic process, model it as a Gaussian white noise, and identify the stochastic differential equation (SDE); this essentially leads to identifying the drift component and diffusion component of a stochastic differential equation. We utilize stochastic calculus to decouple the drift and diffusion part; this allows parallel identification of the two components. Sparse learning and Bayesian statistics, on the other hand, ensures that the governing equation identified is sparse and interpretable; this is achieved by using the spike and slab prior \cite{mitchell1988bayesian, george1997approaches}. The proposed approach has several key features that can be encapsulated into the following points:
\begin{itemize}
	\item First and foremost, unlike other approaches, the proposed approach require only the state measurements; no input measurements are needed.
	\item The proposed approach being Bayesian in nature estimates the posterior distribution. This allows computation of the  epistemic uncertainty (aka predictive uncertainty) due to limited data. This feature is particularly important in design and decision making.
	\item Unlike non-Bayesian approaches, the proposed approach is autonomous in the sense that it doesn't require any human calibration and cross-validation.
	\item The proposed framework falls under the broad umbrella of interpretable machine learning and explainable AI. In other words, the model identified generalizes to unseen environment and out-of-distribution data.
\end{itemize}

Remaining of the paper is structured as follows: in section \ref{sec:methods}, the proposed data-driven framework for learning governing physics without the measurement of input is briefly presented. In section \ref{sec:numerical}, numerical experiments using fully and partially observed output measurements are undertaken for the demonstration of the robustness of the proposed framework. In section \ref{sec:conclusion}, key contributions of the proposed novel framework is summarised and finally the paper is concluded. 

\begin{figure}[htbp!]
	\centering
	\includegraphics[width=\textwidth]{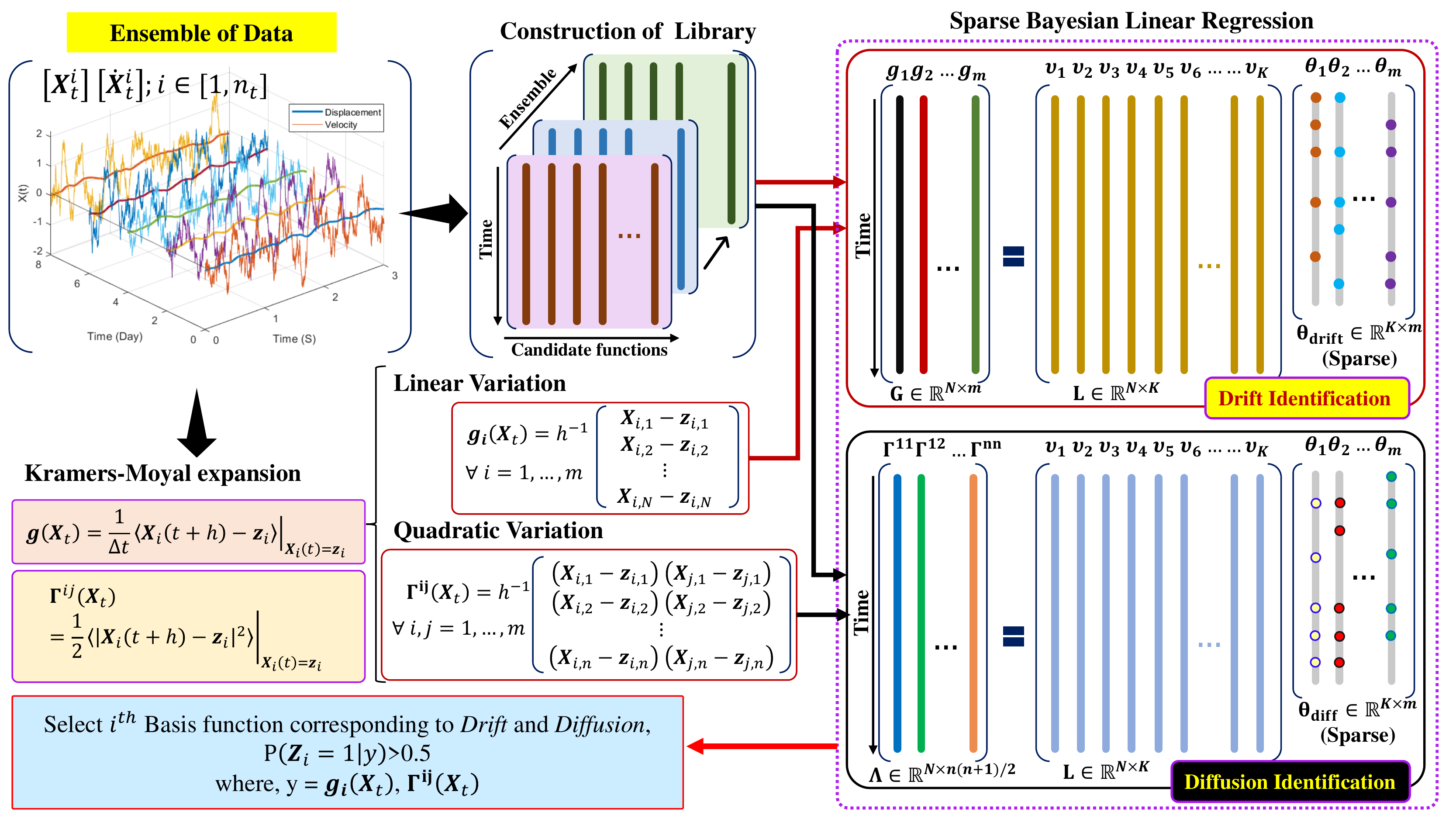}
	\caption{\textbf{Schematic illustration of the proposed Bayesian-SDE framework for discovery of stochastic dynamical systems}. The proposed framework integrates two key ideas for identification of the stochastic systems in terms of the SDEs. The first key step, is the construction of the target vector for the sparse linear regression. For this purpose the proposed framework utilises the Kramers-Moyal expansion to estimate the drift $g({\bm{X}}_t,t)$ and diffusion terms $f({\bm{X}}_t,t)$ in terms of the linear and quadratic variations of the measured sample paths. The linear variation: ${ {\mathop {\lim }_{\Delta t \to 0} \left\langle {{X}(t + \Delta t) - z} \right\rangle }; {X(t) = z}}$ indicates sum of the increments of the sample paths and the quadratic variation: ${ {\mathop {\lim }_{\Delta t \to 0} \left\langle {{{\left| {{X}(t + \Delta t) - z} \right|}^2}} \right\rangle }; {X(t) = z}}$ indicates sum of the square of the increments of the sample paths in limiting sense. The second step, includes formulation of the sparse Bayesian linear regression framework for the above target vector. Given an ensemble of time history of data the framework obtains a sequence of libraries by evaluating the candidate basis functions $\left\{ {{\upsilon _k}({{\bf{X}}_t});\;k = 1,2, \ldots K} \right\}$ over the ensembles. Ensemble mean is taken over all the libraries and the averaged library ${\bf{L}} \in \mathbb{R}^{N \times K}$ is utilized for performing the sparse linear regression. The candidate function in the library are parameterized by a weight vector ${\bm{\theta}}$. To identify each of the drift terms in an $m$-dimensional diffusion process this regression are performed $m$-times. The diffusion terms are not directly recoverable but the discovery of diffusion terms are possible in terms of its covariances. This requires $m(m+1)/2$ number of regression for discovery of all the terms in a covariance matrix. The elements of the weight vector ${\bm{\theta}}$ is assigned a latent variable $Z_k; k=1 \ldots K$ in order to classify the weights into spikes and slabs. Post the Bayesian regression the marginal posterior inclusion probabilities (PIP) $p(Z_k=1|{\bm{Y}})$ is estimated. In the final model only the basis functions whose corresponding marginal PIP values are higher than a threshold value is included. } 
	\label{fig_methodology}
\end{figure}

\section{Discovery of governing physics without input measurement}\label{sec:methods}
The overarching framework of the proposed Bayesian physics discovery framework is presented in the Fig. \ref{fig_methodology}. The proposed framework treats the state measurements from the sensors as stochastic process and applies Kramers-Moyal formula to obtain the target vector for sparse linear regression. In the sparse regression, the library is obtained by taking ensemble mean over all the libraries constructed from the collected data set. Then the proposed framework constructs two sparse regression problem using these libraries and target vectors for identification of the governing physics, which are independent and can be solved simultaneously. The discovery of governing physics here involves identification of the system dynamics and the volatility associated arising due to environmental uncertainty. The system dynamics and the volatility are represented in this framework as drift and diffusion components of an SDE. 

Before formulating the mathematical backbone of the proposed framework let us formalize the problem through following terminologies. Let ${\bf{L}} \in \mathbb{R}^{N \times K}$ be the library of candidate functions, where $K$ is the total number of independent basis functions in the library and $N$ is the sample length. Then the key idea of the sparse Bayesian linear regression is to select $k$ basis functions from the total number of $K$ candidate library functions such that $k \ll K$. If ${\bm{Y}} \in \mathbb{R}^N$ is the target vector then the linear combinations of the identified basis functions should represent the target vector ${\bm{Y}}$ in best possible  coefficients in linear regression are denoted by a weight vector ${\bm{\theta}}$. For the Bayesian model discovery, the condition $k \ll K$ is maintained by assigning certain type of prior distributions on the weight vector ${\bm{\theta}}$, such that they favour the sparsity in the solution. In this context, the SS-prior have shown high shrinkage property due to its sharp spike at zero and a diffused density spanned over a large range of possible parameter value. The Dirac-delta spike concentrates most of the probability mass at zero thus allowing most of the samples to take value zero. On the other hand, the diffused tail distributes a small amount of probability mass over a large range of possible values allowing only very few samples with very high probability to escape the shrinkage. The alternate flavours of the SS-prior constructed from various combination of candidate spike and slab functions are well documented in the literature \cite{mitchell1988bayesian, george1997approaches,o2009review}. In this work, the authors have considered the discontinuous spike and slab prior (DSS-prior) where the spike at zero is modeled as Dirac-delta function and the tail distribution as independent Student’s-t distribution. Next, a sparse regression framework is formulated for SDEs. 

\subsection{Sparse learning of Stochastic differential equations}
Dynamical processes are generally expressed in terms of higher order differential equations. However, in most of the data-driven machine learning techniques the actual system is not observed through its original space but identified in a projected space. The projected space contains all the states of a system which are directly observable. For instance, the second order dynamical systems are often expressed in terms of its displacement and velocity components. Let the projection be realized by a map $\mathcal{T}: \mathbb{R}^{d} \to \mathbb{R}^{m}$ such that $m<d$. If there exists a mapping $\mathcal{T}$, then any higher order system can be reduced to a set of SDEs of the form:
\begin{equation}\label{odegen}
	{\bm{\dot X}} = {\bm{f}}({{\bm{X}}_t},t) + {\bm{g}}({{\bm{X}}_t},t){\bm{\zeta}} (t),
\end{equation}
where ${\bm{f}}({{\bm{X}}_t},t)$ is the deterministic dynamics of the underlying process, ${\bm{g}}({{\bm{X}}_t},t)$ is the volatility associated with the dynamics arising due to the stochastic external input, and ${\bm{\zeta}} (t)$ is the white noise. A more appropriate representation of the above equation can be derived through first order It\^{o} SDEs which arises naturally in non-linear dynamical systems subjected to stochastic excitation such as earthquake, wind force, wave force etc. \cite{calin2015informal,klebaner2005introduction}. In non-linear stochastic dynamical systems the SDEs provide an splendid approach for expressing the behavior of the underlying dynamical system. The SDEs are generally defined in term of its deterministic drift and the additive stochastic diffusion components where both the components are allowed to depend on time and systems states. Let $\left( {\Omega ,\mathcal{F}, P} \right)$ be the probability space and $\{{\mathcal{F}_t},0 \le t \le T)\}$ be the natural filtration constructed from sub $\sigma$-algebras of $\mathcal{F}$. Under the probability space $\left( {\Omega ,\mathcal{F}, P} \right)$ an $m$-dimensional $n$-factor SDE driven by $n$-dimensional Brownian motion \{${\bm{B}}_j(t),j = 1, \ldots n$\} is defined as,  
\begin{equation}\label{sdegen}
\begin{array}{l}
d{\bm{X}}_t = {{\bm f}}\left( {{{\bm{X}}_t},t} \right)dt + \sum\limits_{j = 1}^n {{\bm g}_j\left( {{{\bm{X}}_t},t} \right)} d{{\bf B}_j}\left( t \right); \quad
{\bm X}(t=t_0)={\bm X}_0; \quad t \in [0,T],
\end{array}
\end{equation}
where ${\bm{X}}_t \in {\mathbb{R}^m}$ denotes the ${{\mathcal{F}_t}}$-measurable state vector, ${\bm{f}}\left( {{{\bm{X}}_t},t} \right):{\mathbb{R}^m} \mapsto {\mathbb{R}^m}$ is the drift vector, ${\bm{g}}\left( {{{\bm{X}}_t},t} \right):{\mathbb{R}^m} \mapsto {{\mathbb{R}}^{m \times n}}$ is the diffusion matrix and ${{\bm{B}}_j}\left( t \right) \in {\mathbb{R}^n}$ is the Brownian motion. The white noises are generalized derivative Brownian motion i.e. ${\bm{\zeta}} (t) = \dot {\bm{B}}(t)$; this ensures that there exists a corresponding SDE of Eq. \eqref{sdegen}. In a compact matrix notation, the SDEs are can be expressed as,
\begin{equation}\label{sdeg}
	\begin{array}{l}
		d{{\bm{X}}_t} = {\bm f}\left( {{{\bm {X}}_t},t} \right)dt + {\bm g}\left( {{{\bm {X}}_t},t} \right)d{\bm B}\left( t \right); \quad {\bm X}(t=t_0)={\bm X}_0; \quad t \in [0,T].
	\end{array}
\end{equation}

The solution to Eq. \eqref{sdeg} can be obtained using various stochastic integration schemes \cite{tripura2020ito, kloeden1992higher, tripura2021change, tripura2021generalized}. The discovery of governing physics in terms of an SDE requires the identification of the drift and diffusion components from output measurements. One of the challenges arising in the identification of the SDEs from output data is the non-differentiability of the Brownian motion. In general, the stochastic Brownian motion $\bm B(t)$ is not a well defined mathematical function since it is not differentiable everywhere with respect to the process $\bm X(t)$. However, its continuity is assumed to exists in mean square sense. Formally, if the interval $s\in[0,t]$ is partitioned into $n$-parts as, ${\mathcal{P}_n}(0,t): = \left[ {0 = {s_0} < s_1 <  \ldots  < {s_n} = t} \right]$, then for a  random process $w$, the quadratic variation ${Q_n}(w,t) = \sum\nolimits_i^n {{{\left| {w({s_i}) - w({s_{i - 1}})} \right|}^2}} $ converges to ${Q_n}(w,t) \to Q(w,t)$. From the property of the Brownian motions, it can be found that $Q(w,t)=t$; this states that $B(t)$ has zero finite variation but non-vanishing quadratic variation. On the other hand, the deterministic functions have finite variation but zero quadratic variations. Under these properties, the Kramers-Moyal formula suggests that the drift and diffusion components of an SDE can be extracted from the sample time history in terms of their linear and quadratic variations, respectively. Since the continuity of Brownian motions are mathematically defined in the mean square sense, as a consequence, the diffusion components does not have finite variations but are bounded by their quadratic variations. Thus, the diffusion components, unlike the drifts of an SDE are recoverable only through its covariation terms. The linear and quadratic variations denotes the first and second moments of the sample increments. Leveraging this facts, the drift and diffusion components of an SDE given in Eq. \eqref{sdeg} can be expressed in terms of the sample time history as follows (the simplified detailed derivation is presented in \ref{appenA}):
\begin{equation}\label{kramers}
	\begin{array}{l}
		{{\bm f}_i}\left( {{{\bm {X}}_t},t} \right) = {\left. {\mathop {\lim }\limits_{\Delta t \to 0} \dfrac{1}{{\Delta t}}E\left[ {{X_i}(t + \Delta t) - {\xi_i}} \right]} \right|_{{x_k}(t) = {\xi _k}}}\forall \;k = 1,2, \ldots N\\
		{{\bf \Gamma}_{ij}}\left( {{{\bm {X}}_t},t} \right) = {\left. {\dfrac{1}{2}\mathop {\lim }\limits_{\Delta t \to 0} \dfrac{1}{{\Delta t}}E\left[ {\left| {{X_i}(t + \Delta t) - {\xi_i}} \right|\left| {{X_j}(t + \Delta t) - {\xi_j}} \right|} \right]} \right|_{{x_k}(t) = {\xi_k}}}\forall \;k = 1,2, \ldots N,
	\end{array}
\end{equation}
where ${{\bm f}_i}\left( {{{\bm {X}}_t},t} \right)$ is the $i^{th}$ drift component and ${{\bf{\Gamma}}_{ij}}$ is the $(ij)^{th}$ component of the diffusion covaraince matrix ${\bf{\Gamma}} \in {\mathbb{R}^{n \times n}} := {\bm g}(t,{{\bm{X}}_t}){\bm g}{(t,{{\bm{X}}_t})^T}$. We assume that the analytical form of the drift and diffusion components can be obtained from measurement sample paths in terms of some basis functions. Let ${\upsilon _k}({{\bm{X}}_t})$ represents the various linear and non-linear mathematical functions evaluated on the system states, and $N$ is the length of sample path. Also let ${\bf{L}}^f \in \mathbb{R}^{N \times K}:=[\upsilon _1^f({{\bm{X}}_t}), \ldots, \upsilon _K^f({{\bm{X}}_t})]$ and ${\bf{L}}^g \in \mathbb{R}^{N \times K}:=[\upsilon _1^g({{\bm{X}}_t}), \ldots, \upsilon _K^g({{\bm{X}}_t})]$ are the library of candidate functions constructed from the set $\{ \upsilon _k({{\bm{X}}_t}), k = 1,\ldots,K \}$. Here the candidate functions can contain several functional forms such as polynomial, trigonometric, etc. Then, the $i^{th}$ drift component and the ${ij}^{th}$ term of diffusion covariance matrix can be expressed as linear combination of the library functions as, ${f_i}({{\bm{X}}_t},t) = \sum\nolimits_{k = 1}^K {{\upsilon _k}({{\bm{X}}_t}){\theta _{i,k}}}$; $i=1, \ldots, m$ and ${{\bf{\Gamma }}_{ij}} = \sum\nolimits_{k = 1}^K {{\upsilon _k}({{\bf{X}}_t}){{\bf{\theta }}_{ij,k}}}$; $i,j = 1,2, \ldots m$, briefly,
\begin{equation}
	\begin{array}{l}
		{\bm f}_i({{\bm{X}}_t},t) = \theta _{i_1}^f\upsilon _1^f({{\bm{X}}_t}) +  \ldots  + \theta _{i_k}^f\upsilon _k^f({{\bm{X}}_t}) +  \ldots  + \theta _{i_K}^f\upsilon _K^f({{\bm{X}}_t})\\
		{{\bf \Gamma}_{ij}}\left( {{{\bm {X}}_t},t} \right) = \theta _{{ij}_1}^g\upsilon _1^g({{\bm{X}}_t}) +  \ldots  + \theta _{{ij}_k}^g\upsilon _k^g({{\bm{X}}_t}) +  \ldots  + \theta _{{ij}_K}^g\upsilon _K^g({{\bm{X}}_t}),
	\end{array}
\end{equation}
where $\theta _{i_k}^f$ and $\theta _{{ij}_k}^g$ are the weights associated with the $k^{th}$ basis functions of drift and diffusion covariance components, respectively. 
In general, the libraries ${\bm{L}}^f$ and ${\bm{L}}^g$ can be different but one can construct a library by taking into account of all the basis functions for discovery of both drift and diffusion terms. 
In a general setting, the discovery of the drift terms culminates in the following regression problem,
\begin{equation}
\left[ {\begin{array}{*{20}{c}}
{{{\bm{Y}}_1}}&{{{\bm{Y}}_2}}& \ldots &{{{\bm{Y}}_m}}
\end{array}} \right] = \underbrace {\left[ {\begin{array}{*{20}{c}}
{{\upsilon _1}({X_{1,1}} \ldots {X_{m,1}})}&{{\upsilon _2}({X_{1,1}} \ldots {X_{m,1}})}& \cdots &{{\upsilon _K}({X_{1,1}} \ldots {X_{m,1}})}\\
{{\upsilon _1}({X_{1,2}} \ldots {X_{m,2}})}&{{\upsilon _2}({X_{1,2}} \ldots {X_{m,2}})}& \cdots &{{\upsilon _K}({X_{1,2}} \ldots {X_{m,2}})}\\
 \vdots & \vdots & \ddots & \vdots \\
{{\upsilon _1}({X_{1,N}} \ldots {X_{m,N}})}&{{\upsilon _2}({X_{1,N}} \ldots {X_{m,N}})}& \cdots &{{\upsilon _K}({X_{1,N}} \ldots {X_{m,N}})}
\end{array}} \right]}_{{\bf{L}} \in {^{N \times K}}}\underbrace {\left[ {\begin{array}{*{20}{c}}
{{\theta _{1,1}}}&{{\theta _{2,1}}}& \cdots &{{\theta _{m,1}}}\\
{{\theta _{1,2}}}&{{\theta _{2,2}}}& \cdots &{{\theta _{m,2}}}\\
 \vdots & \vdots & \ddots & \vdots \\
{{\theta _{1,k}}}&{{\theta _{2,k}}}& \cdots &{{\theta _{m,k}}}
\end{array}} \right]}_{{\bf{\theta }} \in {^{K \times m}}}.
\end{equation}

The problem for discovery of diffusion components can also be formulated in the same way as above, thus omitted here for brevity. As a subset of the above regression problems, one can solve the following equations and identify the SDE components independently, 
\begin{subequations}\label{regres_drift1}
\begin{align}
    {{\bm{Y}}_i} &= {\bf{L}}^f{{\bm{\theta }}_i^{f}} + {{\bm{\varepsilon }}_i} \label{subeq:regres_drift}\\
    {{\bm{Y}}_{ij}} &= {\bf{L}}^g{{\bm{\theta }}_{ij}^{g}} + {{\bm{\eta }}_{ij}}, \label{subeq:regres_diffusion}
\end{align}
\end{subequations}
where ${{\bm{Y}}_i}$ and ${{\bm{Y}}_{ij}}$ are the target vectors of the sparse regression problem associated with $i^{th}$-drift component and $(ij)^{th}$-diffusion covariance term, respectively, and,  ${{\bm{\varepsilon }}_i}$ and ${{\bm{\eta }}_{ij}}$ are the corresponding residual error vectors. For the discovery of the $i^{th}$ drift component in Eq. \eqref{subeq:regres_drift}, the target vector is defined as, ${{\bm{Y}}_i} = {{\Delta t}^{-1} \left[ {\left( {{X_{i,1}} - {\xi _{i,1}}} \right) \ldots \left( {{X_{i,N}} - {\xi _{i,N}}} \right)} \right]^T}$, and in the formulation of the sparse regression problem for diffusion components in Eq. \eqref{subeq:regres_diffusion}, the target vector is set as, ${{\bm{Y}}_{ij}} = {{\Delta t}^{-1} \left[ {\left( {{X_{i,1}} - {\xi _{i,1}}} \right)\left( {{X_{j,1}} - {\xi _{j,1}}} \right) \ldots \left( {{X_{i,N}} - {\xi _{i,N}}} \right)\left( {{X_{j,N}} - {\xi _{j,N}}} \right)} \right]^T}$. The weights vectors for $i^{th}$ drift component and ${ij}^{th}$ term of diffusion covariance matrix are given as, ${{\bm{\theta }}_i^{f}} = {\left[ {{\theta _{1_1}}},{{\theta _{1_2}}}, \ldots ,{{\theta _{1_K}}} \right]^T}$, and ${\bm{\theta }}_{ij}^g = {\left[ {\theta _{{ij}_1}},{\theta _{{ij}_2}}, \ldots ,{\theta _{{ij}_K}} \right]^T}$, respectively. The straightforward application of the Gibbs sampling mentioned in the Eqs. \eqref{gibbs_theta} $\to$ \eqref{gibbs_p0} can be performed to solve the regression problems in Eq. \ref{regres_drift1}. Noting that a $m$-dimensional process has $m$ numbers of drift components it requires $m$-independent regressions to identify all the drifts. Further it can be noted that in a $m$-dimensional process with $m \times n$-diffusion matrix the diffusion covariance matrix ${\bf \Gamma}$ has a dimension $m \times m$. However, due to the symmetry of the covariance matrix, it requires ${{m(m + 1)}}/{2}$ numbers of independent regressions to be performed to completely identify the diffusion space.

\subsection{Discovery of SDE by sparse Bayesian regression}\label{section21}
Since Eq. \eqref{subeq:regres_drift} and \eqref{subeq:regres_diffusion} reflects the same sparse regression problem to be solved, for the brevity they can be represented using the following one-dimensional regression equation:
\begin{equation}\label{regression1}
	{\bm{Y}} = {\bf{L}}{\bm{\theta}} + \epsilon,
\end{equation}
where ${\bm{Y}} \in \mathbb{R}^N$ denotes the $N$-dimensional target vector, ${\bf{L}}$ denotes the library of candidate functions, ${\bm{\theta}}$ is the weight vector, and $\epsilon \in \mathbb{R}^N$ is the residual error vector representing the model mismatch error. In the Bayesian inference, given the output measurements the aim is to find the posterior distribution of the weight vector ${\bm \theta}$. For estimating the posterior distribution of the weight vector ${\bm{\theta}}$, one can apply the Bayes formula in Eq. \eqref{regression1}, which yields,
\begin{equation}\label{bayes1}
	P\left( {\bm \theta |{\bm{Y}}} \right) = \frac{P\left( \bm \theta \right){P\left( {{\bm{Y}}|{\bm{\theta}} } \right)}}{{P\left( {\bm{Y}} \right)}}.
\end{equation}
Modeling the mismatch error $\epsilon$ as i.i.d Gaussian random variable with zero mean and variance $\sigma^2$, the likelihood function is written as,
\begin{equation}\label{liklihood}
	{\bm{Y}}|{\bm{\theta }},{\sigma ^2} \sim \mathcal{N}\left( {{\bf{L}}{\bm{\theta}} ,{\sigma ^2}{{\bf{I}}_{N \times N}}} \right),
\end{equation}
where ${\bf{I}}_{N \times N}$ denotes the ${N \times N}$ identity matrix. The discovery of governing physics demands that the resulting model should be interpretable i.e. the solution of weight vector ${\bm \theta}$ should contain only few terms in the final model. In this work, the discontinuous spike and slab (DSS) distributions are used for promotion of sparsity in the solution. The equation discovery using DSS-prior requires the classification of weights of the basis functions into either of the spike and slab component. This is done by introducing a latent indicator variable ${\bm{Z}}=\left[ Z_1, \ldots, Z_K \right]$ for each of the component $\theta_k; k = 1, \ldots, K$ of the weight vector ${\bm{\theta}}$. The latent indicator variable $Z_k$ is assigned a value 1 if the weight corresponds to the slab component, otherwise, it takes a value 0 when the weight belongs to spike component. It is to be noted that the components of the weight vector that belongs to spike does not contribute to discovery of governing physics. Therefore, let us consider a vector ${\bm{\theta}}_r \in \mathbb{R}^r: \{r \ll K\}$, composed from the elements of the weight vector ${\bm{\theta}}$ for which $Z_k=1$. Denoting ${\bm{\theta}}_r$ as the weight vector containing only those variables from ${\bm{\theta}}$ for which $Z_k=1$, the DSS-prior is defined as \cite{nayek2021spike,mitchell1988bayesian},
\begin{equation}\label{dss}
	p\left( {{\bm{\theta }}|{\bm{Z}}} \right) = {p_{slab}}({\theta _r})\prod\limits_{k,{Z_k} = 0} {{p_{spike}}({\theta _k})},
\end{equation}
where the spike and slab distributions are defined as, ${p_{spike}}({\theta _k}) = {\delta _0}$ and ${p_{slab}}({{\bm \theta} _r}) = \mathcal{N} \left( {{\bf{0}},{\sigma ^2}{\vartheta _s}{{\bf{R}}_{0,r}}} \right)$ with ${{\bf{R}}_{0,r}} = {{\bf{I}}_{r \times r}}$. It can be observed that the spike and slab distributions are assumed to be independent. To increase the effectiveness and faster convergence to the sparse solution the noise variance ${\sigma ^2}$ and the slab variance ${\vartheta _s}$ are treated as random variables. The noise variance $\sigma^2$ is assigned the Inverse-gamma distribution with the hyperparameters ${\alpha _\sigma }$ and ${\beta _\sigma }$, the the slab variance $\vartheta_s$ is assigned the Inverse-gamma prior with the hyperparameters ${\alpha _\vartheta }$ and ${\beta _\vartheta }$, the latent variables $Z_k$ takes a value from the set $\{0, 1\}$, thus each of the element in the latent vector is assigned the Bernoulli prior with common hyperparameter ${p_0}$. The hyperparameter ${p_0}$ is simulated from the Beta prior with the hyperparameters ${\alpha _p }$ and ${\beta _p }$. This can be summarized as,
\begin{align}
	& p\left({\vartheta _s} \right) = IG\left( {{\alpha _\vartheta },{\beta _\vartheta }} \right) \label{hyper1} \\ 
	& p\left({Z_k}|{p_0} \right)  = Bern\left( {{p_0}} \right);k = 1 \ldots K \label{hyper2} \\ 
	& p\left( {p_0} \right) = Beta\left( {{\alpha _p},{\beta _p}} \right) \label{hyper3} \\ 
	& p\left({\sigma ^2} \right) = IG\left( {{\alpha _\sigma },{\beta _\sigma }} \right) \label{hyper4} 
\end{align}

\begin{figure}[ht]
	\centering
	\includegraphics[width=0.3\textwidth]{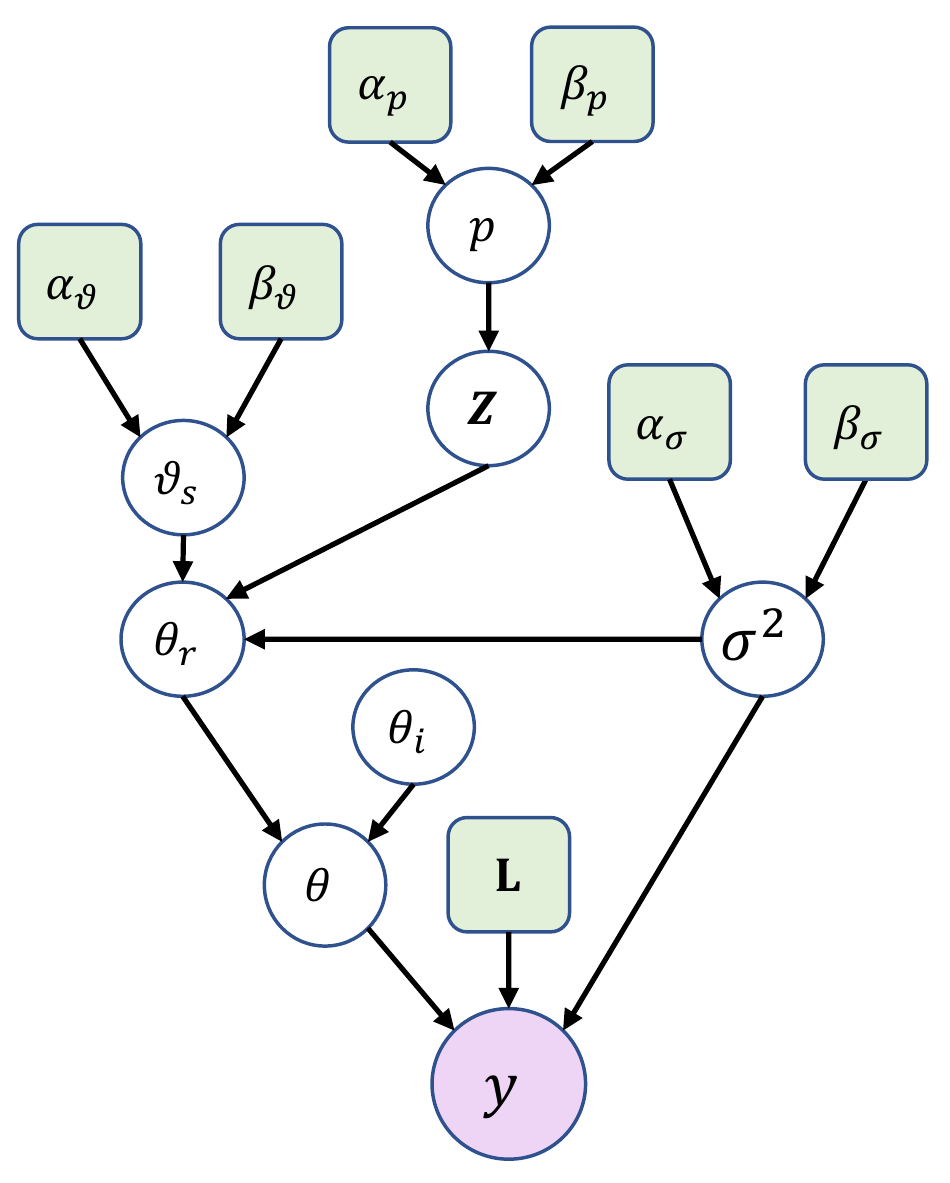}
	\caption{\textbf{Hierarchical Bayesian network of the discontinuous spike and slab model for sparse linear regression}. The variables in the green square boxes indicate the deterministic parameters and the variables in the white circles represents random variables. The term ${\bf{L}}$ represents the library of candidate functions, and the parameters ${\alpha _\vartheta }$, ${\beta _\vartheta }$, ${\alpha _p }$, ${\beta _p }$, ${\alpha _\sigma }$, and ${\beta _\sigma }$ indicates the hyperparameters of the priors of the hierarchical DSS model. Here $p$, $\vartheta_s$ and $\sigma^2$ scalar valued, and, the the variables ${\bm{Z}}$ and ${\bm{\theta}}$ are vector valued variables.}
	\label{fig_graph}
\end{figure}

With all the random variables, we construct a bigger Bayesian hierarchical model as shown in Fig. \ref{fig_graph}, where the hyperparameters ${\alpha _\vartheta }$, ${\beta _\vartheta }$, ${\alpha _p }$, ${\beta _p }$, ${\alpha _\sigma }$, and ${\beta _\sigma }$ are provided as a deterministic constants in the hierarchical model. From the DAG structure in Fig. \ref{fig_graph}, the joint distribution of the random variables ${\bm{\theta }}$, ${\bm{Z}}$, ${\vartheta _s}$, ${\sigma ^2}$, ${p_0}|{\bm{Y}}$ are obtained from the Fig. \ref{fig_graph} as,
\begin{equation}\label{joint1}
	\begin{array}{ll}
		p\left( {{\bm{\theta }},{\bm{Z}},{\vartheta _s},{\sigma ^2},{p_0}|{\bm{Y}}} \right) &= \dfrac{{p\left( {{\bm{Y}}|{\bm{\theta }},{\sigma ^2}} \right)p\left( {{\bm{\theta }}|{\bm{Z}},{\vartheta _s},{\sigma ^2}} \right)p\left( {{\bm{Z}}|{p_0}} \right)p\left( {{\vartheta _s}} \right)p\left( {{\sigma ^2}} \right)p\left( {{p_0}} \right)}}{{p\left( {\bm{Y}} \right)}}\\
		& \propto p\left( {{\bm{Y}}|{\bm{\theta }},{\sigma ^2}} \right)p\left( {{\bm{\theta }}|{\bm{Z}},{\vartheta _s},{\sigma ^2}} \right)p\left( {{\bm{Z}}|{p_0}} \right)p\left( {{\vartheta _s}} \right)p\left( {{\sigma ^2}} \right)p\left( {{p_0}} \right),
	\end{array}
\end{equation}
where $p\left( {{\bm{\theta }},{\bm{Z}},{\vartheta _s},{\sigma ^2},{p_0}|{\bm{Y}}} \right)$ denotes the joint distribution of the random variables, $p\left( {{\bm{Y}}|{\bm{\theta }},{\sigma ^2}} \right)$ denotes the likelihood function, $p\left( {{\bm{\theta }}|{\bm{Z}},{\vartheta _s},{\sigma ^2}} \right)$ is the prior distribution for the weight vector ${\bm{\theta}}$, $p\left( {{\bm{Z}}|{p_0}} \right)$ is the prior distribution for the latent vector ${\bm{Z}}$, $p\left( {{\vartheta _s}} \right)$ is the prior distribution for the slab variance ${\vartheta}_s$, $p\left( {{\sigma ^2}} \right)$ is the prior distribution for the noise variance, $p\left( {{p_0}} \right)$ is the prior distribution for the success probability $p_0$ and ${p\left( {\bm{Y}} \right)}$ is the marginal likelihood. Direct sampling from the joint distribution function is intractable in this case due to the spike and slab distribution. Thus, Gibbs sampling technique is used to draw the random samples from the joint distribution \cite{casella1992explaining}. For the Gibbs sampling, the conditional distributions of the random variables are derived in \ref{appenB}. The weights corresponding to the spike distribution does not contribute to the selection of the correct basis functions. Thus, only the weights corresponding to $Z_k \ne 0$ i.e. the ${\bm{\theta}}_r$ vector is sampled using the Gibbs sampling. Referring to the Eqs. \eqref{hyper1}, \eqref{hyper2}, \eqref{hyper3}, and \eqref{hyper4}, the sequence of the random variables ${{\bm{\theta }}^{(0)}},{\sigma ^{2(0)}},\vartheta _s^{(0)},p_0^{(0)},{{\bm{Z}}^{(0)}}, \ldots ,{{\bm{\theta }}^{(1)}},{\sigma ^{2(1)}},\vartheta _s^{(1)},p_0^{(1)},{{\bm{Z}}^{(1)}}, \ldots $, using the Gibbs sampling technique is obtained by following steps,
\begin{enumerate}
    \item The weight vector ${\bm{\theta}}_r^{(i)}$ is sampled from the Gaussian distribution with mean ${{\bm{\mu}} _\theta }$ and variance ${{\bf{\Sigma}} _\theta }$ as,
    \begin{equation}\label{gibbs_theta}
        {{\bm{\theta }}_r^{(i)}}|{\bm{Y}},{\vartheta _s^{(i)}},{\sigma ^{2(i)}} \sim \mathcal N\left( {{{\bm{\mu}} _\theta ^{(i)}},{{\bf{\Sigma}} _\theta ^{(i)}}} \right),
    \end{equation}
    where the mean and covariance is defined as, ${{\bm{\mu}} _\theta^{(i)} } = {\bf{\Sigma}}_\theta^{(i)} {\bf{L}}_r^{(i)T}{\bm{Y}}$ and ${{\bf{\Sigma}} _\theta ^{(i)}} = {\sigma ^{2(i)}}{\left( {{\bf{L}}_r^{(i)T}{{\bf{L}}_r^{(i)}} + \vartheta _s^{ {(i)}- 1}{\bf{R}}_{0,r}^{{(i)} - 1}} \right)^{ - 1}}$, respectively.
    
    \item The latent variable $Z_k^{(i+1)}$ is assigned the values from the set, $\left\{0,1 \right\}$ by using the Bernoulli distribution as,
    \begin{equation}\label{gibbs_z}
        {Z_k^{(i+1)}}|{\bm{Y}},{\vartheta _s^{(i)}},{p_0^{(i)}} \sim Bern\left( {{u_k}} \right),
    \end{equation}
    where ${u_k} = \dfrac{{{p_0}}}{{{p_0} + \lambda \left( {1 - {p_0}} \right)}}$ and $\lambda  = \dfrac{{p\left( {{\bm{Y}}|{Z_k^{(i)}} = 0,{{\bm{Z}}_{ - k}^{(i)}},{\vartheta _s^{(i)}}} \right)}}{{p\left( {{\bm{Y}}|{Z_k^{(i)}} = 1,{{\bm{Z}}_{ - k}^{(i)}},{\vartheta _s^{(i)}}} \right)}}$. Here, ${{\bm{Z}}_{ - k}^{(i)}} \in \mathbb{R}^{K-1}$ denotes the latent variable vector ${\bm{Z}}$ consisting of all the elements except the $k^{th}$ component. The probability that the $k^{th}$ latent variable $Z_k^{(i)}$ takes a value 0 or 1 is estimated as follows,
    \begin{equation}
    \begin{array}{ll}
      p\left( {{\bm{Y}}|{\bm{Z}}^{(i)},{\vartheta _s^{(i)}}} \right) & = \dfrac{{\Gamma \left( {{\alpha _\sigma } + \dfrac{N}{2}} \right)\beta _\sigma ^{{\alpha _\sigma }}}}{{\Gamma \left( {{\alpha _\sigma }} \right){{\left( {2\pi } \right)}^{\dfrac{N}{2}}}{{\left( {{\beta _\sigma } + \dfrac{1}{2}{{\bm{Y}}^T}{\bm{Y}}} \right)}^{\left( {{\alpha _\sigma } + \dfrac{N}{2}} \right)}}}}; \quad \text{when all $\left\{ {Z_k^{(i)}}{: k = 1, \ldots ,K} \right\} = 0$}\\
      
     & = \dfrac{{\Gamma \left( {{\alpha _\sigma } + \dfrac{N}{2}} \right)\beta _\sigma ^{{\alpha _\sigma }}{{\left( {\left| {{\bf{R}}_{0,r}^{{(i)} - 1}} \right|\left| {{{\bf{\Sigma}} _\theta ^{(i)}}} \right|} \right)}^{\dfrac{1}{2}}}}}{{\Gamma \left( {{\alpha _\sigma }} \right){{\left( {2\pi } \right)}^{\dfrac{N}{2}}}\vartheta _s^{\dfrac{{{h_z}}}{2}}{{\left( {{\beta _\sigma } + \dfrac{1}{2}{{\bm{Y}}^T}\left( {{{\bf{I}}_{N \times N}} - {{\bf{L}}_r^{(i)}}{{\bf{\Sigma}} _\theta ^{(i)}}{\bf{L}}_r^{(i)T}} \right){\bm{Y}}} \right)}^{\left( {{\alpha _\sigma } + \dfrac{N}{2}} \right)}}}}; \quad \text{otherwise.}
        \end{array}
    \end{equation}
    
    \item The noise variance ${\sigma ^{2(i+1)}}$ is simulated from the Inverse-gamma distribution as,
    \begin{equation}\label{gibbs_sigma}
        {\sigma ^{2(i+1)}}|{\bm{Y}},{\bm{Z}^{(i+1)}},{\vartheta _s^{(i)}} \sim IG\left( {{\alpha _\sigma } + \dfrac{N}{2},{\beta _\sigma } + \dfrac{1}{2}\left( {{{\bm{Y}}^T}{\bm{Y}} - {\bm{\mu}} _\theta^{(i)T} {\bf{\Sigma}} _\theta ^{{(i)} - 1}{{\bm{\mu}} _\theta ^{(i)}}} \right)} \right).
    \end{equation}
    
    \item The slab variance ${\vartheta _s^{(i+1)}}$ is sampled from the Inverse-gamma distribution as,
    \begin{equation}\label{gibbs_slab}
        {\vartheta _s^{(i+1)}}|{\bm{\theta} ^{(i)}},{\bm{Z}}^{(i+1)},{\sigma ^{2(i+1)}} \sim IG\left( {{\alpha _\vartheta } + \dfrac{{{h_z}}}{2},{\beta _\vartheta } + \dfrac{1}{{2{\sigma ^2}}}{\bm{\theta }}_r^{(i)T}{\bf{R}}_{0,r}^{{(i)} - 1}{{\bm{\theta }} _r^{(i)}}} \right).
    \end{equation}
    
    \item The success rate ${p_0^{(i+1)}}$ is sampled from the Beta distribution as,
    \begin{equation}\label{gibbs_p0}
        {p_0^{(i+1)}}|{\bm{Z}}^{(i+1)} \sim Beta\left( {{\alpha _p} + {h_z},{\beta _p} + K - {h_z}} \right),
    \end{equation}
    where ${h_z} = \sum\nolimits_{k = 1}^K {{Z_k}^{(i+1)}}$.
    
    \item The weight vector ${\bm{\theta}}_r^{(i+1)}$ is updated using the step 1.
\end{enumerate}
This MCMC is performed for a total of ${N_M}$ simulations out of which initial 1000 samples are discarded as the burn-in samples. Let ${N_s}$ denote the number of MCMC required to achieve the stationary distribution after the burn-in samples are discarded. Then the marginal posterior inclusion probability (PIP):= $p\left( {{Z_k} = 1|{\bm{Y}}} \right)$ for each of the $K$ basis functions can be estimated by taking mean over the Gibbs samples for each of the $k^{th}$ latent vector ${Z_k}$ \cite{nayek2021spike} as,
\begin{equation}\label{mpip}
p\left( {{Z_k} = 1|{\bm{Y}}} \right) \approx \dfrac{1}{{{N_s}}}\sum\limits_{j = 1}^{{N_s}} {Z_k^j} ;{\rm{ }}k = 1, \ldots ,K.
\end{equation}
The basis functions whose corresponding PIP values are more than 0.5 i.e. $p\left( {{Z_k} = 1|{\bm{Y}}} \right) > 0.5$ are included in the final model of discovered equation. The PIP value greater than 0.5 indicates that the selected basis functions are observed more than half of the times in the MCMC simulations. Higher PIP value suggests that in case of unseen scenarios, the corresponding basis functions are highly likely to occur in the data representation target vector. The mean and covariance of the weight vector gives the expected value and standard deviation of the system parameters. The standard deviation provides the confidence interval of the parameters which can be used to design the lower and upper bounds of a random variable in an unseen environment. A pseudo code for the proposed framework, is provided in the Algorithm \ref{algvecvec}.

\begin{algorithm}
	\caption{Pseudo code of the proposed Bayesian framework for discovery of governing physics of stochastic non-linear systems}\label{algvecvec}
	\begin{algorithmic}[1]
	\Require{Sample paths: ${\bf{X}}(t) \in \mathbb{R}^{N \times m}$, hyperparameters: $\alpha_p$, $\beta_p$, $\alpha_\sigma$, $\beta_\sigma$, $\alpha_\vartheta$, $\beta_\vartheta$, $p_0^{(0)}$, $\vartheta_s^{(0)}$}
	    \State{Estimate the linear and quadratic variation vectors.} \Comment{Eq. \eqref{kramers}}
	    \State{For drift obtain the target vectors $\bm{Y}_i$ and the library ${\bf{L}}$ using the candidate basis functions.}  \Comment{Eq. \eqref{subeq:regres_drift}}
	    \State{For diffusion obtain the target vectors $\bm{Y}_{ij}$ and the library ${\bf{L}}$}. \Comment{Eq. \eqref{subeq:regres_diffusion}}
	    \State{Estimate the initial variance of noise from residual variance: $\sigma^{2,(0)}$ = Var(${\bf{L}}{\bm{\theta}}-{\bm{Y}}$)}
	    \State{Estimate the initial latent vector $\bm{Z}^{(0)}$=$\left[Z_1^{(0)}, Z_2^{(0)}, \ldots, Z_K^{(0)} \right]$ such that MSE$({\bf{L}}{\bm{\theta}}-{\bm{Y}})$ is minimum.}
	    \State{Find $\theta^{(0)}$, ${\bm{\mu}}_\theta$ and ${\bf{\Sigma}}_\theta$} \Comment{Eq. \eqref{gibbs_theta}}
		\For {$i$ = $1, \ldots, {N_M}$}	
		\State Update the latent variable vector ${\bm{Z}}^{(i)}$. \label{MCMC1} \Comment{Eq. \eqref{gibbs_z}}
		\State Update the noise variance $\sigma^{2(i)}$. \Comment{Eq. \eqref{gibbs_sigma}}
		\State Update the slab variance $\vartheta _s^{(i)}$. \Comment{Eq. \eqref{gibbs_slab}}
		\State Update the success rate $p_0^{(i)}$. \Comment{Eq. \eqref{gibbs_p0}}
		\State Update the weight vector ${\bm{\theta}}^{(i)}$. \label{MCMCn} \Comment{Eq. \eqref{gibbs_theta}} 
		\State{Repeat steps \ref{MCMC1}$\to$\ref{MCMCn}}
		\EndFor
	\State{Discard the burn-in MCMC samples}
	\State{Estimate the marginal PIP values $p(Z_k=1|{\bm{Y}})$} \Comment{Eq. \eqref{mpip}}
	\State{Include the basis functions with higher PIP values}
	\State{Estimate the expected value ${\rm E}[{\theta_k}]$ and standard deviation ${\sigma}[{\theta_k}]$ of the system parameters}
	\Ensure{$\{ {\bm{\theta}}_{k};k=1 \ldots {K} \}$, $\{ \vartheta_k ({\bm{X}}_j); j=1 \ldots {\bar{m}} \}$ }, where $K$ and $\bar{m}$ are the number of library functions and process states, respectively.
	\end{algorithmic}
\end{algorithm}

The Gibbs sampler is initialised with the following initial values of the hyperparameters: $p_0^{(0)}$=0.1, $\vartheta^{(0)}$=10, and $\sigma ^{2(0)}$ is set equal to the residual variance from ordinary least-squares regression. The initial vector of binary latent variables ${\bm{Z}}^{(0)}$ is computed by setting ${\bm{Z}}^{(0)}=[Z_1, \ldots, Z_K]$ to zero and
then activating the components of ${\bm{Z}}$ that reduce the mean-squared error between the training data and the obtained model from ordinary least-squares. For this purpose a forward followed by a backward search algorithm is devised, where the backward search iterates through the activated components of initial latent vector in similar fashion to forward search. Given all the other parameters, the initial value of $\theta^{(0)}$ is obtained from the Eq. \eqref{joint1}. For the commencement of the algorithm, the deterministic prior parameters are set to the following values: $a_p$=0.1 and $b_p$=1 for the Beta prior on $p_0$, $a_v$=0.5 and $b_v$=0.5 for inverse-Gamma prior on slab variance, and, $a_\sigma$=$10^4$ and $b_\sigma$=$10^4$ for inverse-Gamma prior on measurement noise. A Markov chains with 3000 samples is used for Gibbs sampling. The first 1000 samples are discarded as burn-in, and the remaining 2000 samples are used for posterior computation.

For all the demonstrations in this work, the data are simulated using the Euler-Maruyama (EM) scheme at a frequency of 1000Hz using the parameters listed in the Table \ref{table_param}. The noise in the measurements is modeled as $N$-dimensional sequence of zero-mean Gaussian white noise with a standard deviation equal to 5\% of the standard deviation of the simulated quantities. In this work, the dictionary ${\bf{L}} \in \mathbb{R}^{N \times K}$ is constructed from 5 types of mathematical functions, each function representing a mapping of the $m$-dimensional state vector $\bm{X} = \left\{ X_1, X_2, \ldots X_m \right\}$:
\begin{equation}
	{\bf{L}}({\bm{X}}) = \left[ {\begin{array}{*{20}{c}}
			{\bf{1}}&{{P^1}({\bm{X}})}&{{P^2}({\bm{X}})}& \ldots &{{P^\mathcal{P}}({\bm{X}})}&{{\mathop{\rm sgn}} ({\bm{X}})}&{\left| {\bm{X}} \right|}&{{\bm{X}} \otimes \left| {\bm{X}} \right|} 
	\end{array}} \right].
\end{equation}
Here ${\bf{1}} \in \mathbb{R}^{N}$ denotes the $N$-dimensional vector of 1, ${P^\mathcal{P}}({\bm{X}}) \in \mathbb{R}^{N \times m}$ denotes the set of terms present in the multinomial expansion ${({X_1} + {X_2} +  \ldots  + {X_m})^\mathcal{P}}$, $sgn({\bm{X}}) \in \mathbb{R}^{N \times m}$ represents the signum functions of the form $sgn({X_i})$ $\forall i = 1 \ldots m$, ${\left| {\bm{X}} \right|} \in \mathbb{R}^{N \times m}$ denotes the absolute mapping of the states: ${\left| {X_i} \right|}$ $\forall i = 1 \ldots m$. The tensor product term ${{\bm{X}} \otimes \left| {\bm{X}} \right|} \in \mathbb{R}^{N \times {2m}}$ represents the set of functions: ${X_i \otimes \left| X_j \right|}$ $\forall {i,j} = 1 \ldots m$. For the study, the length of the polynomial $\mathcal{P}$ is chosen as 6. The cardinality of the library is found as $K = (1 + \left| {{{({X_1} + {X_2} +  \ldots  + {X_m})}^n}} \right| + 4m)$, where the number of terms in the multinomial expansion can be found as, $\left| {{{({X_1} + {X_2} +  \ldots  + {X_m})}^n}} \right| = {}^{n + m - 1}{C_{m - 1}}$. The complete architecture of the propose framework is illustrated in the Fig. \ref{fig_methodology}.
\begin{table}[ht]
	\centering
	\begin{tabular}{llll}
		\hline
		\textbf{Simulated systems} & \textbf{Drift parameters} & & \multirow{2}{*}{\textbf{Diffusion parameters}} \\ \cline{2-3}
		& \textbf{Linear} & \textbf{Non-linear} & \\ 
		\hline
		Black-Scholes & $\lambda$ = 2 & - & $\mu$ = 1 \\
		\hline
		Duffing-Van der pol & $m$ = 1, $k$ = $2 \times 10^3$, $c$ = 2 & $\alpha$ = $10^5$ & $\sigma$ = 10 \\
		\hline
		2-DOF non-linear & $m_1$ = $m_2$ = 1, $k_1$ = $4 \times 10^3$ & $\mu$ = 1, $g$ = 9.81 & $\sigma_1$ = 10, $\sigma_2$ = 10 \\
		& $k_2$ = $2 \times 10^3$, $c_1$ = $c_2$ = 2 && \\
		\hline
		Bouc-Wen & $m$ = 1, $k$ = $1 \times 10^4$, $c$ = 20, $\lambda$ = 0.5  & $A_1$ = $A_2$ = 0.5, $A_3$ = 1, $\bar{n}$ = 3 & $\sigma$ = 2 \\
		\hline
	\end{tabular}
	\caption{Simulation parameters of the systems. }
	\label{table_param}
\end{table}

\section{Discovery of governing physics of example problems}\label{sec:numerical}
\begin{figure}[htbp!]
	\centering
	\includegraphics[width=\textwidth]{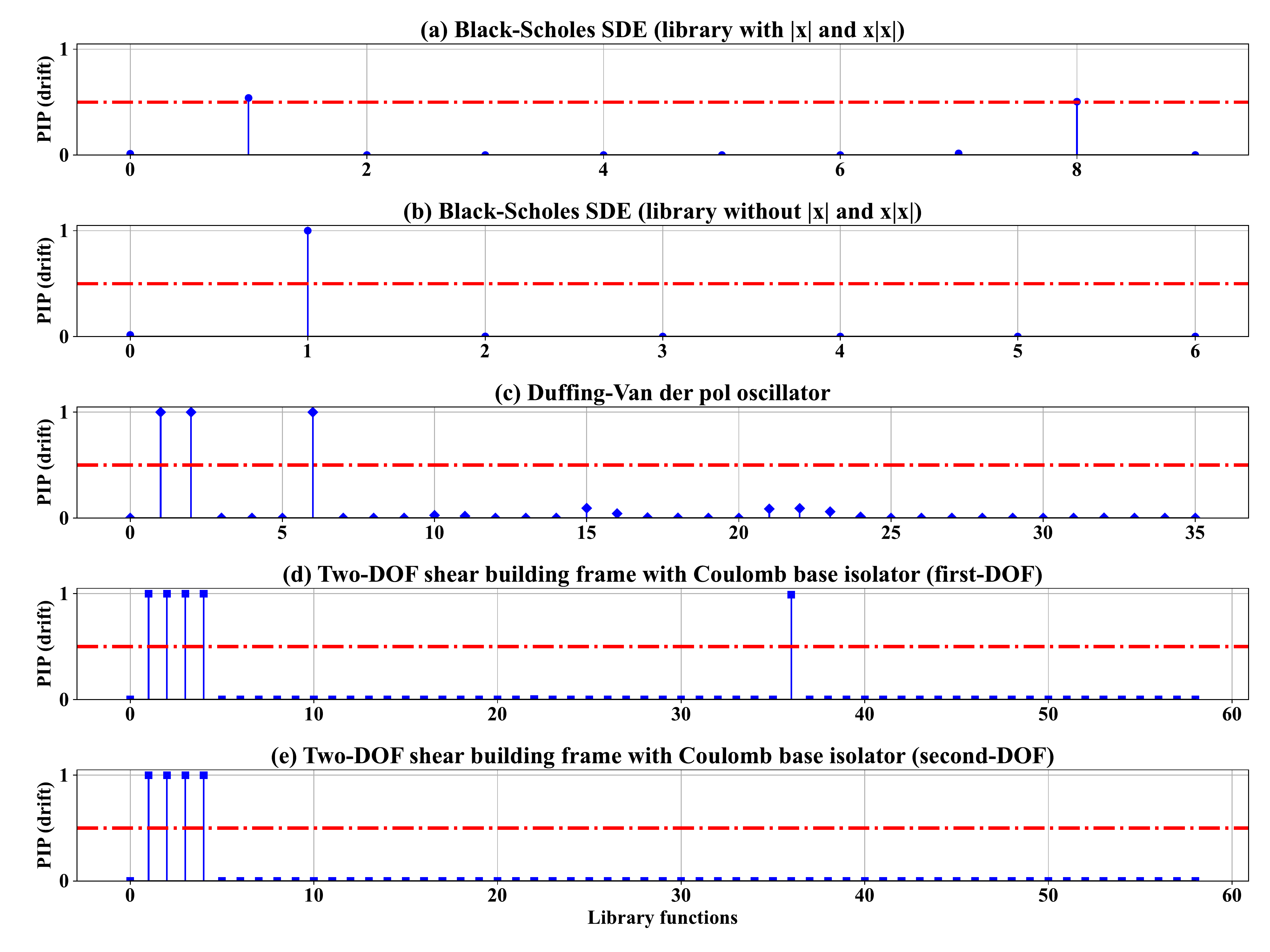}
	\caption{\textbf{Discovery of the drift components of the governing physics from data corrupted with 5\% white Gaussian noise based on marginal posterior inclusion probability (PIP), P$(Z_{k}=1|\bm{Y})$}. (a) Black-Scholes SDE: the library ${\bf{L}} \in \mathbb{R}^{N \times 9}$ consists 9 candidate basis functions out of which the basis functions $x$ and $|x|$ are identified based on the criteria, PIP>0.5. (b) Black-Scholes SDE: the library ${\bf{L}} \in \mathbb{R}^{N \times 6}$ is constructed as a collection of 6 basis functions.  The correct drift component $X(t)$ is accurately identified with the almost sure probability ${\rm{P}}(Z_1=1)=1$. (c) Duffing-Van der pol oscillator: the library ${\bf{L}} \in \mathbb{R}^{N \times 36}$ has 36 candidate basis functions. The drift component is correctly identified as, $\upsilon(1)=X_1$, $\upsilon(2)=X_2$ and $\upsilon(6)=X_1^3$. (d) Two-DOF shear building (drift component of first DOF): the library ${\bf{L}} \in \mathbb{R}^{N \times 58}$ consists a total of 58 candidate basis functions, out of which 5 basis are selected for discovery of first drift component as, $\upsilon(1)=Y_1$, $\upsilon(2)=Y_2$, $\upsilon(3)=Y_3$, $\upsilon(4)=Y_4$ and $\upsilon(36)=sgn(Y_2)$. (e) Two-DOF shear building (drift component of first DOF): from the library ${\bf{L}} \in \mathbb{R}^{N \times 58}$ the basis functions are discovered as, $\upsilon(1)=Y_1$, $\upsilon(2)=Y_2$, $\upsilon(3)=Y_3$, $\upsilon(4)=Y_4$. The basis functions for both the drift components are selected with almost full probability ${\rm{P}}(Z_k=1|\bm{Y})$.}
	\label{figdrift}
\end{figure}

\begin{figure}[htbp!]
	\centering
	\includegraphics[width=\textwidth]{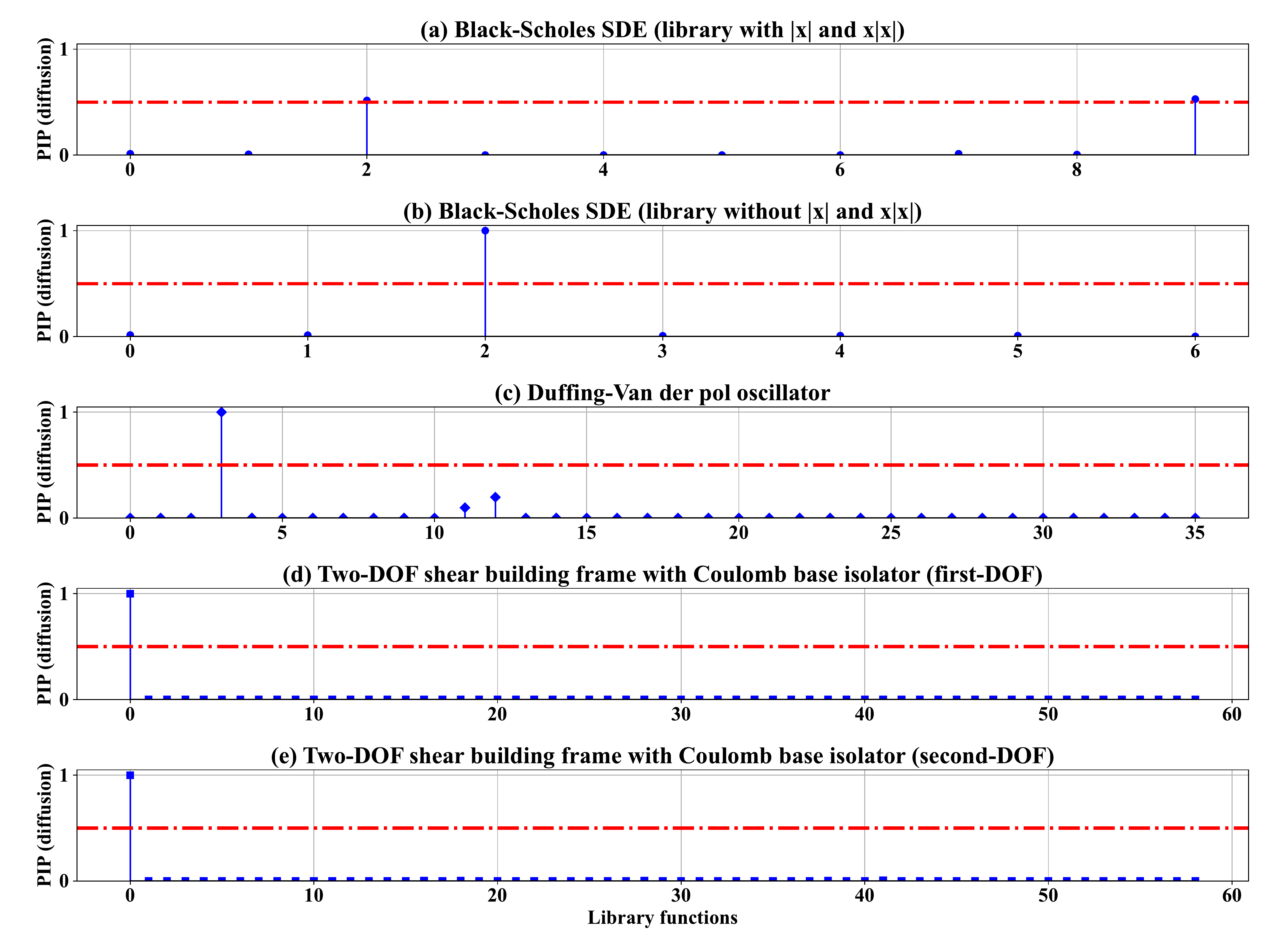}
	\caption{\textbf{Discovery of the diffusion components of the governing physics from data corrupted with 5\% white Gaussian noise based on marginal posterior inclusion probability (PIP), P$(Z_{k}=1|\bm{Y})$}. (a) Black-Scholes SDE: the weight vector $\theta \in \mathbb{R}^{9}$ for the library ${\bf{L}} \in \mathbb{R}^{N \times 9}$ is shown. The covariance of the diffusion component is identified as a combination of the terms $X^2(t)$ and $X|X|$. (b) Black-Scholes SDE: the weight vector $\theta \in \mathbb{R}^{6}$ has 6 elements out of which the the correct covariance term $X^2(t)$ is discovered with almost sure probability ${\rm{P}}(Z_2=1)=1$. (c) Duffing-Van der pol oscillator: there are 36 elements in the weight vector $\theta \in \mathbb{R}^{36}$. The diffusion term is discovered correctly as $X_1^2(t)$ with almost sure probability ${\rm{P}}(Z_2=1)=1$. (d)  Two-DOF shear building (drift component of first DOF): the weight vector $\theta \in \mathbb{R}^{58}$ has 58 elements corresponding to the 58 basis functions. The first diffusion component is accurately identified as $X_1^2(t)$ with almost sure probability ${\rm{P}}(Z_2=1)=1$. (e) Two-DOF shear building (drift component of first DOF): the second diffusion component $X_2^2(t)$ is accurately identified with almost sure probability ${\rm{P}}(Z_2=1)=1$.}
	\label{figdiff}
\end{figure}

In this section, the efficacy and robustness of the proposed Bayesian physics discovery framework is observed on a variety of representative stochastic non-linear dynamical systems. The examples taken includes: (a) Black–Scholes SDE, (b) Duffing Van-der pol oscillator, and (c) Two degree-of-freedom (DOF) base isolated shear building. For the equation discovery, it is assumed that the input forces are not measurable and only the noisy measurements of the displacements are available for physics discovery. The velocity component is obtained from the displacement vector through numerical differentiation such as fourth order central finite difference formula. Additionally a case study using Bouc-Wen oscillator is undertaken, where it is assumed that the hysteresis measurement is also not available. Therefore the proposed framework treats it as a partially observed system and tries to estimate the effect of the unobserved state without explicitly considering it in to the library of candidate functions. The results of the case studies undertaken in this work are presented in Figs. \ref{figdrift}, \ref{figdiff} and Table \ref{table_summery}. These results demonstrate sufficient accuracy in the discovered physics and robustness of the proposed framework to learn governing law from limited and noisy displacement measurements only (without any input measurements).

\begin{figure}[htbp!]
	\centering
	\includegraphics[width=\textwidth]{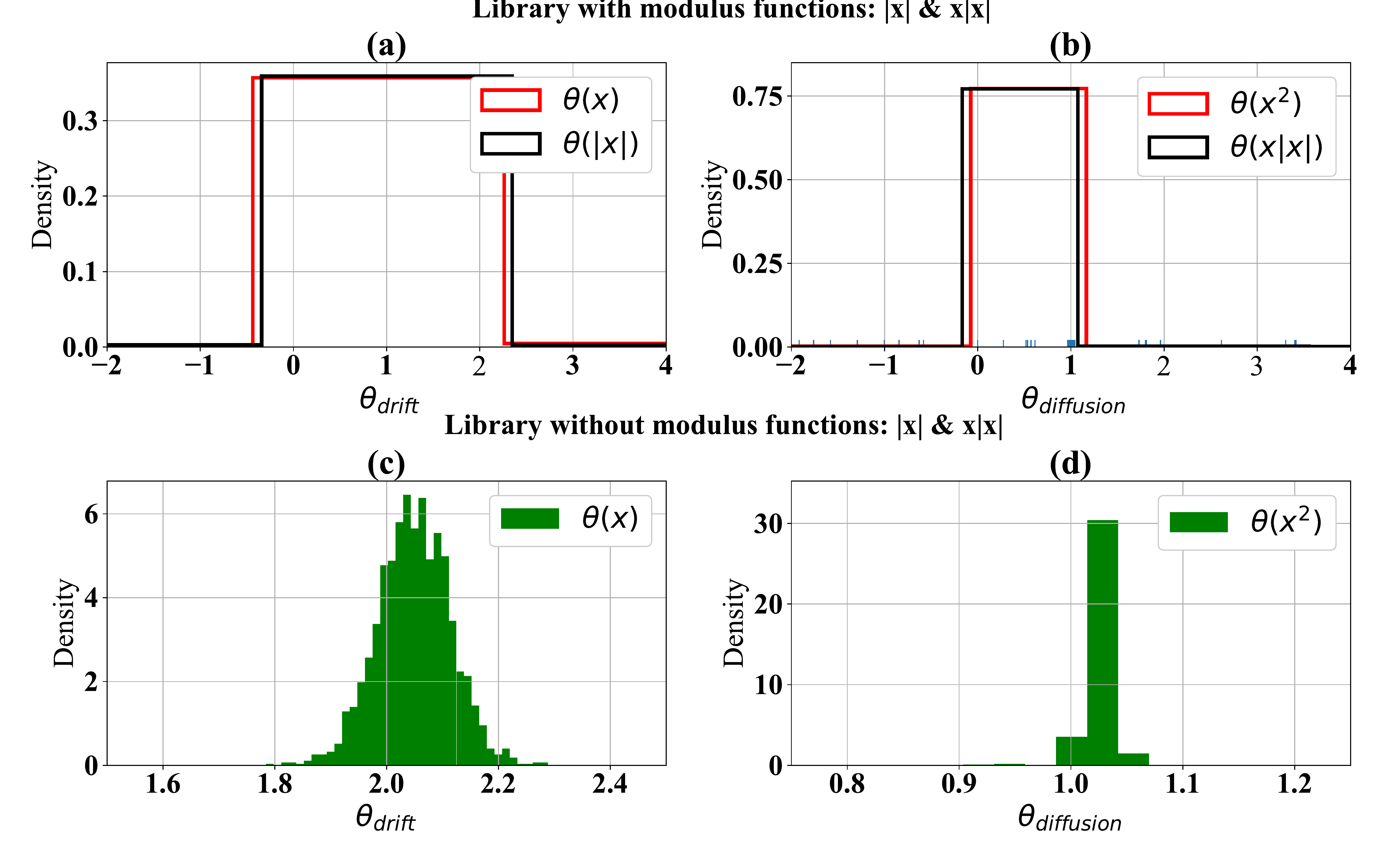}
	\caption{\textbf{Black-Scholes SDE: posterior distributions of the weights of discovered basis functions}. (a) The posterior distributions of $\theta_{X}$ and $\theta_{|X|}$. Posterior of both emulates the uniform distribution, whose means lie around the value 1. As a combination the basis $\theta_{X}$ and $\theta_{|X|}$ identifies the drift. (b) The posterior distributions of $\theta_{X^2}$ and $\theta_{X|X|}$. The posteriors emulate uniform distribution. The mean gives a value around 0.5. Thus, as a combination they discovered the diffusion. (c) The posterior of weight $\theta_{X}$. The mean is obtained as 2.04, which exactly identifies the drift with almost sure probability ${\rm{P}}(Z_1=1)=1$. (d) The posterior of weight $\theta_{X^2}$ whose mean provides the value of the parameters $\mu$ as 1.02.}
	\label{figblack2}
\end{figure}
\begin{table}[ht!]
	\centering
	\begin{tabular}{ll}
		\hline
		\textbf{Correct SDEs}: &\\ \cline{1-1}
		Black-Scholes: & $dX(t) = 2X(t)dt + X(t)dB(t)$\\
		Duff.-Van der pol: & $d{X_2}(t) =  - \left( {1000{X_1}(t) + 2{X_2}(t) + 100000X_1^3(t)} \right)dt + 10X(t)dB(t)$\\
		2-DOF non-linear: & $\left\{ \begin{array}{l}
			d{Y_2}(t) = \left( { - 6000{Y_1}(t) - 4{Y_2}(t) - 9.81{\mathop{\rm sgn}} \left( {{Y_2}(t)} \right) + 2000{Y_3}(t) + 2{Y_4}(t)} \right)dt + 10d{B_1}(t)\\
			d{Y_4}(t) = \left( {2000{Y_1}(t) + 2{Y_2}(t) - 2000{Y_3}(t) - 2{Y_4}(t)} \right)dt + 10d{B_2}(t)
		\end{array} \right.$\\
		\hline 
		\textbf{Identified SDEs}: &\\ \cline{1-1}
		Black-Scholes: & $\left\{ \begin{array}{l}
			dX(t) = 2.05X(t)dt + 1.02X(t)dB(t)\\
			dX(t) = \left( {0.47X(t) + 0.53\left| {X(t)} \right|} \right)dt + \left( {1.09{X^2}(t) + 0.95X(t)\left| {X(t)} \right|} \right)dB(t)
		\end{array} \right.$ \\
		Duff.-Van der pol: & $d{X_2}(t) =  - \left( {1000.02{X_1}(t) + 1.99{X_2}(t) + 99885.10X_1^3(t)} \right)dt + 10.31X(t)dB(t)$\\
		2-DOF non-linear: & $\left\{ \begin{array}{ll}
			d{Y_2}(t) =& \left( { - 6000.30{Y_1}(t) - 3.97{Y_2}(t) - 9.89{\mathop{\rm sgn}} \left( {{Y_2}(t)} \right) + 1999.73{Y_3}(t) + 2.09{Y_4}(t)} \right)dt \\ 
			&+ 10.03d{B_1}(t)\\
			d{Y_4}(t) =& \left( {2000.65{Y_1}(t) + 1.99{Y_2}(t) - 1999.06{Y_3}(t) - 1.99{Y_4}(t)} \right)dt + 10.11d{B_2}(t)
		\end{array} \right.$ \\
		\hline
	\end{tabular}
	\caption{\textbf{Summary of the results of Bayesian equation discovery of stochastic dynamical systems}. The parameters of the identified systems denote the expected value of the weights after discarding the burn-in samples. Here, the first 1000 samples are taken as burn-in samples and therefore discarded for obtaining the final stationary distribution.}
	\label{table_summery}
\end{table}

\subsection{Black–Scholes SDE}
A Black–Scholes SDE (formulated as geometric Brownian motion) is a continuous-time stochastic process where the logarithm of the randomly varying quantity follows a Brownian motion. The Black-Scholes SDE is frequently used in stock market for modelling of the evolution of stock price of an underlying asset \cite{oksendal2013stochastic}. The Black–Scholes SDE has the drift $f(t,X_t) = \lambda X$ and diffusion $g(t,X) = \mu X$ with $\lambda > 0$ and $\mu > 0$ being the real constants. Towards this, the Black–Scholes SDE is defined as a geometric Brownian motion as follows:

\begin{equation}\label{eq_BS}
    \dfrac{{dX(t)}}{{X(t)}} = \lambda dt + \mu dB(t)
    \quad
{ X}(t=t_0)={ X}_0; \quad t \in [0,T],
\end{equation}
where $B=\{B(t); t\ge 0\}$ is the Brownian motion. Data is generated by simulating Eq. \eqref{eq_BS} and then corrupting it with Gaussian white noise. 
Identifying the diffusion term as, $g({\bf{X}}_t,t)=\mu X(n)$, the variance of the diffusion is obtained: $\Gamma= \mu^2 X^2(n)$. The results of this case study are plotted in the Figs. \ref{figdrift}(a) and \ref{figdiff}(a). Fig. \ref{figdrift}(a) depicts that there are two identified basis function $X$ and $|X|$. However, from the Eq. \eqref{eq_BS}, it is obvious that only one function $X$ should have been identified as the driving function of the Black-Scholes SDE. To understand this discrepancy one needs to consider that the evolution of the random variable $X(t)$ in the Black-Scholes SDE follows Geometric Brownian motion (GBM) and GBMs always assume positive values (e.g. real stock price). Since the characteristics of both the functions $X$ and $|X|$ in this case are same, the algorithm identifies both the functions with almost equal probability. A similar phenomenon is observed in case of the identification of the diffusion term too. Figs. \ref{figdiff}(b) depicts almost equal contribution from the functions $X^2$ and $X|X|$, however, only the term $X^2$ should have been actually identified. Upon considering the corresponding parameter values of the basis functions $X$ and $|X|$ for the drift component, and, $X^2$ and $X|X|$ for the diffusion component, the Black-Scholes SDE in Eq. \eqref{eq_BS} can be easily simulated for any unseen environmental conditions. 

To verify that this is not a limitation of the proposed scheme, this numerical demonstration is repeated in the Figs. \ref{figdrift}(b) and \ref{figdiff}(b) without considering the functions $|X|$ and $X|X|$ in the library. The parameters of the basis functions for the drift and diffusion components are portrayed in Figs. \ref{figblack2}(c) and \ref{figblack2}(d), respectively. The results clearly show that the proposed scheme is able to identify the basis functions along with their associated parameters $\lambda$ and $\mu$ without any significant error. As a consequence of the above case study, one can ask a question about what basis functions are to be considered in the library. One of the possibilities could be to visualize the time series data and then take scientific and engineering judgements. For example, if the time history data shows no zero crossing then one can opt against considering the functions which shares similar properties (for e.g. $|X|$ and $X$, and $X|X|$ and $X^2$). Besides one can also choose to consider due to the fact that the similar terms will have equal probability of occurrence and as a combination will demonstrate the observed phenomenon. However, for future prediction the model will be applicable to process depicting no zero crossing only.

\begin{figure}[htbp!]
	\centering
	\includegraphics[width=\textwidth]{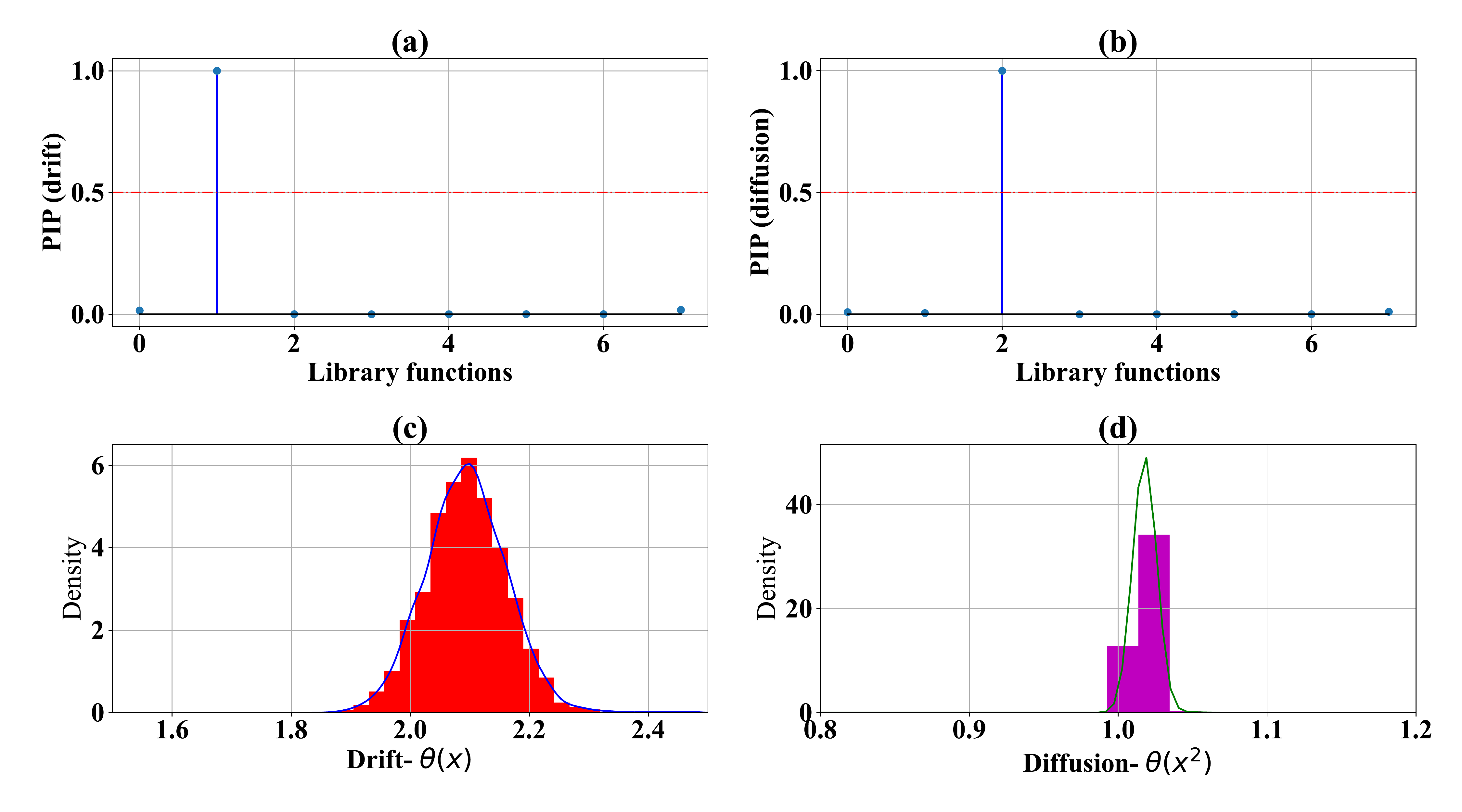}
	\caption{\textbf{Black-Scholes equation: Posterior distributions of the weights of the selected basis functions}. (a) The drift component $X$ of the Black-Scholes equation is identified. (b) The variance term $X^2$ of the diffusion component is identified. (c) Posterior distribution of the weight $\theta (X)$ which identifies the expected value of the parameter $\lambda$. The expected value of $\lambda$ is identified from the mean of the distribution as approximately 2.09. (d) The posterior distribution of the weight $\theta (X^2)$. The mean of the distribution represents the expected value of parameter $\mu$, which is found as 1.01. }
	\label{figBS_t15}
\end{figure}

Finally, to illustrate the robustness of the proposed approach and simulate a realistic scenario, we consider a case where the data is generated using strong Taylor 1.5 stochastic integration scheme. The results of the study is provided in the Fig. \ref{figBS_t15}. In the Figs. \ref{figBS_t15}(a) and (b), it is evident that the proposed framework is able to correctly identify the drift and diffusion components of the Black-Scholes equation. The subplots (c) and (d) further shows the expected values of the parameters of the identified basis functions, which matches almost exactly with the parameters listed in Table. \ref{table_param}. This shows that despite the lower order formulation, the proposed framework is robust and well equipped.

\subsection{Duffing-Van der pol oscillator}
\begin{figure}[htbp!]
	\centering
	\includegraphics[width=0.8\textwidth]{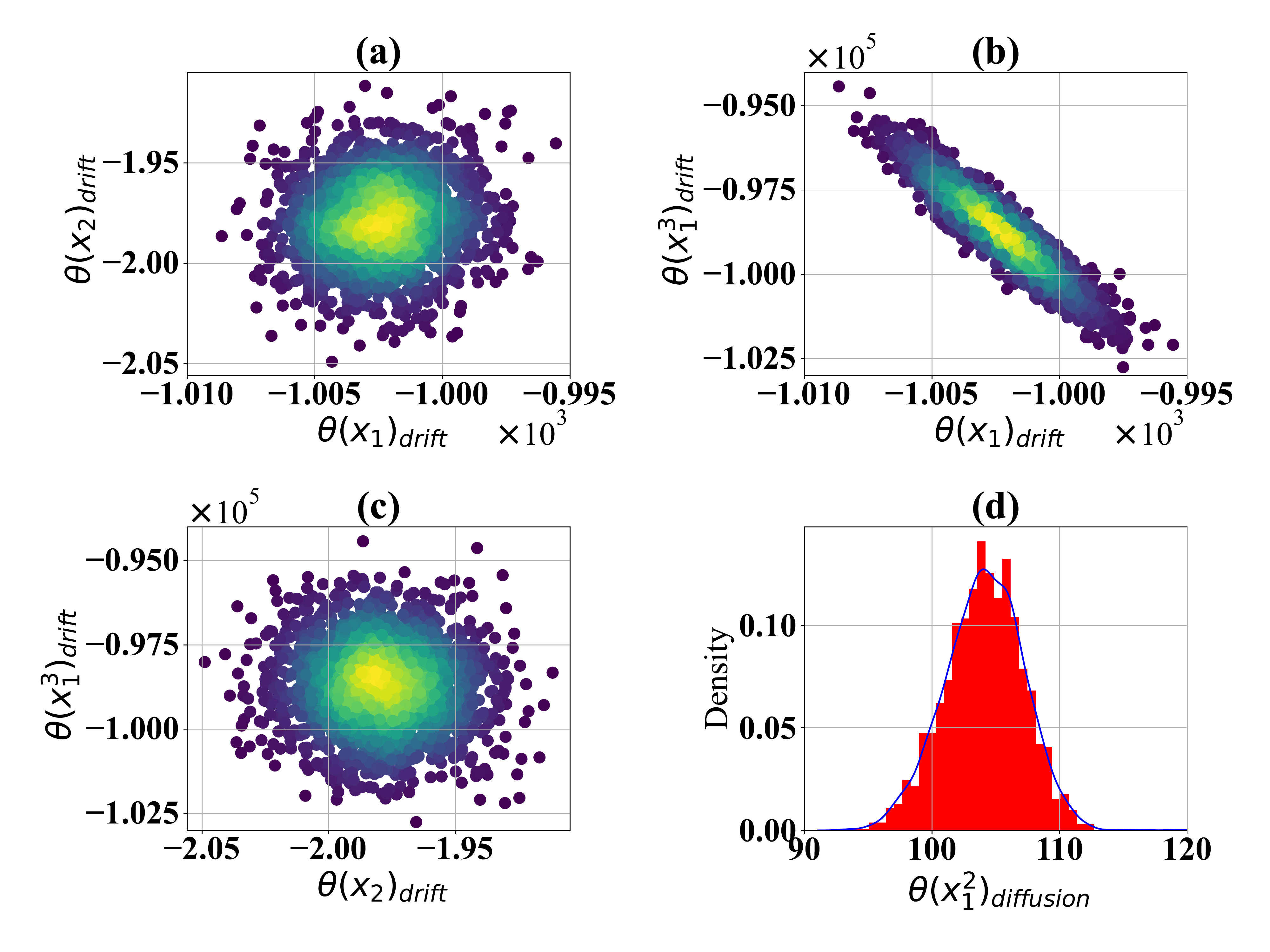}
	\caption{\textbf{Duffing-Van der pol oscillator: Posterior distributions of the weights of the selected basis functions}. (a) Joint posterior distribution of $\theta(X_1)$ and $\theta(X_2)$. The yellow region indicates the mean of the joint distribution. This weights $\theta(X_1)$ and $\theta(X_2)$ represents the parameters $k$ and $c$. The expected values of the identified weights are, ${\rm{E}}(\theta (X_1))=1000.2$ and ${\rm{E}}(\theta (X_2))=1.99$. (b) Joint posterior distribution of $\theta(X_1)$ and $\theta(X_1^3)$. The expected values of the weights are ${\rm{E}}(\theta (X_1))=1000.2$ and ${\rm{E}}(\theta (X_1^3))=99885.1$. (c)	Joint posterior distribution of $\theta(X_2)$ and $\theta(X_1^3)$. he expected values of the weights are ${\rm{E}}(\theta (X_2))=1.99$ and ${\rm{E}}(\theta (X_1^3))=99885.1$. (d) The histogram of the diffusion parameter. The mean value of the histogram is 106.47 which represents the term $g^2(\bm{X}_t,t)$. Upon conversion the parameters $\sigma/m$ can be obtained as 10.29. }
	\label{figdvp_param}
\end{figure}

The second order non-linear hardening Duffing-Van der pol oscillator is considered, in this section. The Duffing-Van der pol oscillator draws its physical relevance from the models in flow-induced structural vibration problems. The Duffing-Van der pol oscillator has a cubic dissipating force and it is driven by multiplicative white noise. Towards this, the equation of motion of the undertaken system is expressed as:
\begin{equation}
    m\ddot X(t) + c\dot X(t) + kX(t) + \alpha {X^3}(t) = \sigma X(t)\dot B(t)
    \quad
{\bm X}(t=t_0)={\bm X}_0; \quad t \in [0,T],
\end{equation}
where $m$, $c$, and $k$ are the mass, damping, and stiffness parameters of the oscillator. Here, $\alpha$ is a real-valued parameter associated with the cubic non-linearity, $\sigma \ge 0$ is the strength of the multiplicative white noise, and $B=\{B(t); t\ge 0 \}$ is the Brownian motion. The time derivative of the Brownian $\dot B(t)$ here represents the white noise. With a statespace transformation $X = {X_1}$, and $\dot X = {X_2}$, where $X_1$ and $X_2$ represents the displacement and velocity, the corresponding first order It\^{o}-stochastic SDEs for the dynamical system are derived as,
\begin{equation}
    \left[ {\begin{array}{*{20}{c}}
{d{X_1}(t)}\\
{d{X_2}(t)}
\end{array}} \right] = \left[ {\begin{array}{*{20}{c}}
{{X_2}(t)}\\
{\dfrac{{ - 1}}{m}\left( {k{X_1}(t) + c{X_2}(t) + \alpha X_1^3(t)} \right)}
\end{array}} \right]dt + \left[ {\begin{array}{*{20}{c}}
0\\
{\dfrac{\sigma }{m}X(t)}
\end{array}} \right]dB(t).
\end{equation}
For estimating the drift and diffusion terms only, the evolution of second variable ${X_2}(t)$ is used to construct the target linear and quadratic variation vectors from the sample paths using the Kramers-Moyal formulae. Here, the variance of the diffusion term is identified as, $\Gamma = ({\sigma^2}X^2(t))/{m^2}$. The results for the basis function selection are displayed in Figs. \ref{figdrift}(c) and \ref{figdiff}(c), respectively, while their associated parameters are plotted in Fig. \ref{figdvp_param}. From the Figs. \ref{figdrift}(c) and \ref{figdiff}(c), it is evident that the proposed framework is able to accurately discover the basis functions for the drift and diffusion terms as, $\left\{X_1(t), X_2(t), X_1^3(t) \right\}$ and $\left\{ X_1(t), X_2(t) \right\}$, respectively.

The pairwise joint posterior distributions of weights associated with the discovered basis functions are presented in Fig. \ref{figdvp_param}. In the subplots (a), (b), and (c), the weights of discovered drift functions are shown. It is evident in this figures that the actual parameters associated with the drift component of DVP oscillator as listed in Table \ref{table_param} is correctly identified with very small relative error. The subplot (d) depicts the posterior distribution of the identified diffusion basis function. The mean value of the posterior distribution represents the term ${g^2}({{\bm{X}}_t},t)$, which in this case can be noticed as $g({{\bm{X}}_t},t)$ = $X(t)$. The mean value here is obtained as approximately 106.47. By performing the square root operation over the mean, one can approximately get the diffusion value $\sigma = 10.31$. These results verify that the proposed scheme can be successfully applied to discover the governing physics of non-linear oscillators when subjected to parametric excitation effectively.

\subsection{Two story base isolated Shear Building}
\begin{figure}[t]
	\centering
	\includegraphics[width=\textwidth]{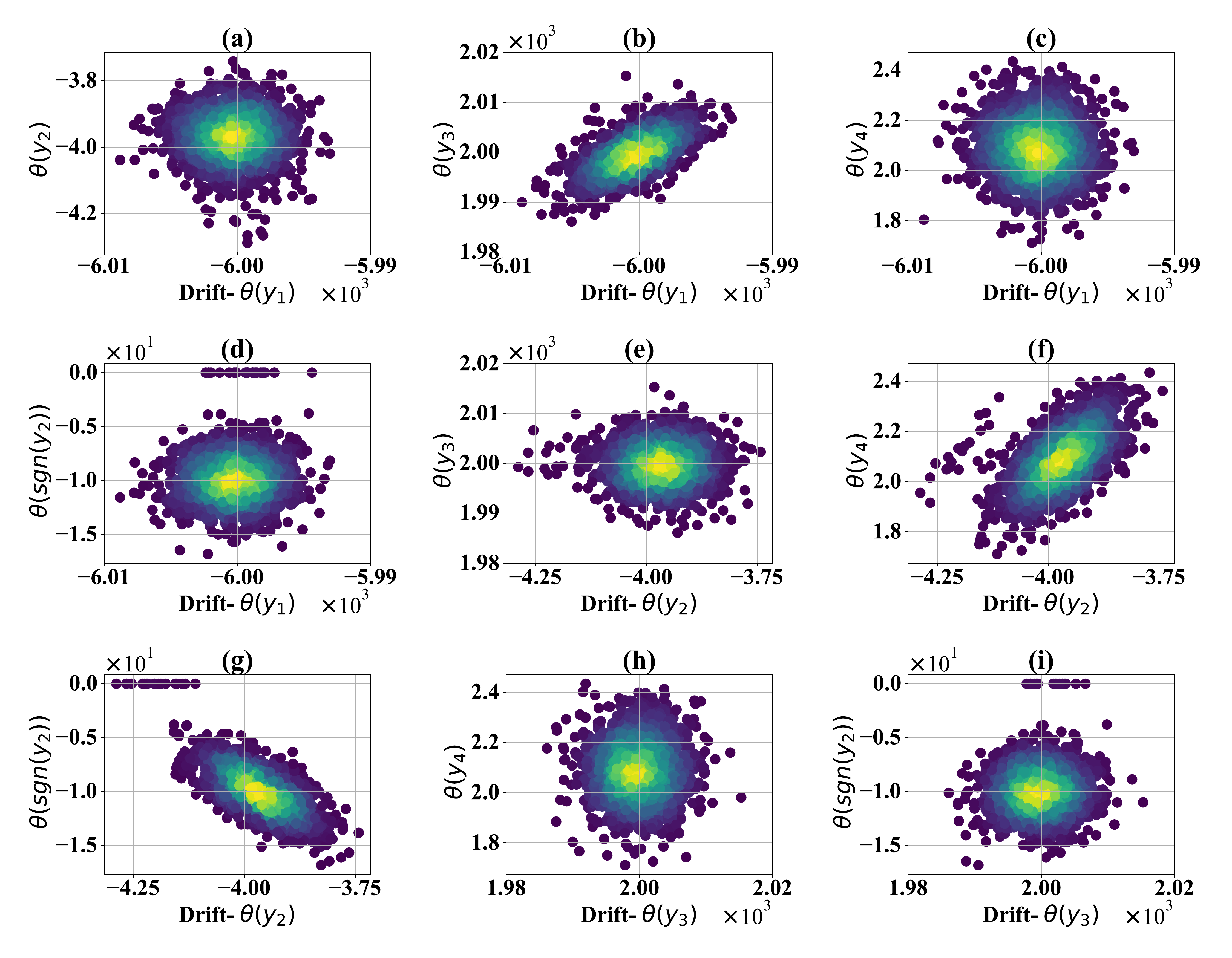}
	\caption{\textbf{Non-linear two-DOF oscillator (first drift component): Pairwise joint posterior distributions of the parameters $\theta_k$ of the selected basis functions}. (a) Joint posterior of the weights $\theta({Y_1})$ and $\theta(Y_2)$. The yellow color denotes the mean value of the joint distribution. The expected values of the parameters are identified as ${\rm{E}}(\theta (Y_1))$= -6000.30 and ${\rm{E}}(\theta (Y_2))$= -3.97. (b) Joint posterior of the weights $\theta_{Y_1}$ and $\theta_{Y_3}$. The expected values of the parameters are identified as ${\rm{E}}(\theta (Y_1))$= -6000.30 and ${\rm{E}}(\theta (Y_3))$= 1999.73. (c) Joint posterior of the weights $\theta_{Y_1}$ and $\theta_{\upsilon_1}$. The expected values of the parameters are identified as ${\rm{E}}(\theta (Y_1))$= -6000.30 and ${\rm{E}}(\theta (Y_4))$= 2.09. (d) Joint posterior of the weights $\theta_{Y_1}$ and $\theta_{sgn(Y_2)}$. The expected values of the parameters are identified as ${\rm{E}}(\theta (Y_1))$= -6000.30 and ${\rm{E}}(\theta (sgn(Y_2)))$= -9.89. (e) Joint posterior of the weights $\theta_{Y_2}$ and $\theta_{Y_3}$. The expected values of the parameters are identified as ${\rm{E}}(\theta (Y_2))$= -3.97 and ${\rm{E}}(\theta (Y_3))$= 1999.73. (f) Joint posterior of the weights $\theta_{Y_2}$ and $\theta_{Y_4}$. The expected values of the parameters are identified as ${\rm{E}}(\theta (Y_2))$= -3.97 and ${\rm{E}}(\theta (Y_4))$= 2.09. (g) Joint posterior of the weights $\theta_{Y_2}$ and $\theta_{sgn(Y_2)}$. The expected values of the parameters are identified as ${\rm{E}}(\theta (Y_2))$= 1999.73 and ${\rm{E}}(\theta (sgn(Y_2)))$= -9.89. (h) Joint posterior of the weights $\theta_{Y_3}$ and $\theta_{Y_4}$. The expected values of the parameters are identified as ${\rm{E}}(\theta (Y_3))$= 1999.73 and ${\rm{E}}(\theta (Y_4))$= 2.09. (i) Joint posterior of the weights $\theta_{Y_3}$ and $\theta_{sgn(Y_2)}$. The expected values of the parameters are identified as ${\rm{E}}(\theta (Y_3))$= 1999.73 and ${\rm{E}}(\theta (sgn(Y_2)))$= -9.89. The weights $\{\theta_{Y_1}, \theta_{Y_2}, \theta_{Y_3}, \theta_{Y_4}, \theta_{sgn(Y_2)}\}$ corresponds to the basis functions $\{\upsilon_1, \upsilon_2, \upsilon_3, \upsilon_4, \upsilon_36 \}$. They represents the relative value of the parameters $\{k_1, k_2, k_3, k_4, \mu mg \}$. The negative values arises due to the coupling of the systems states. }
	\label{fig2dof_param_drift1}
\end{figure}

\begin{figure}[t]
	\centering
	\includegraphics[width=\textwidth]{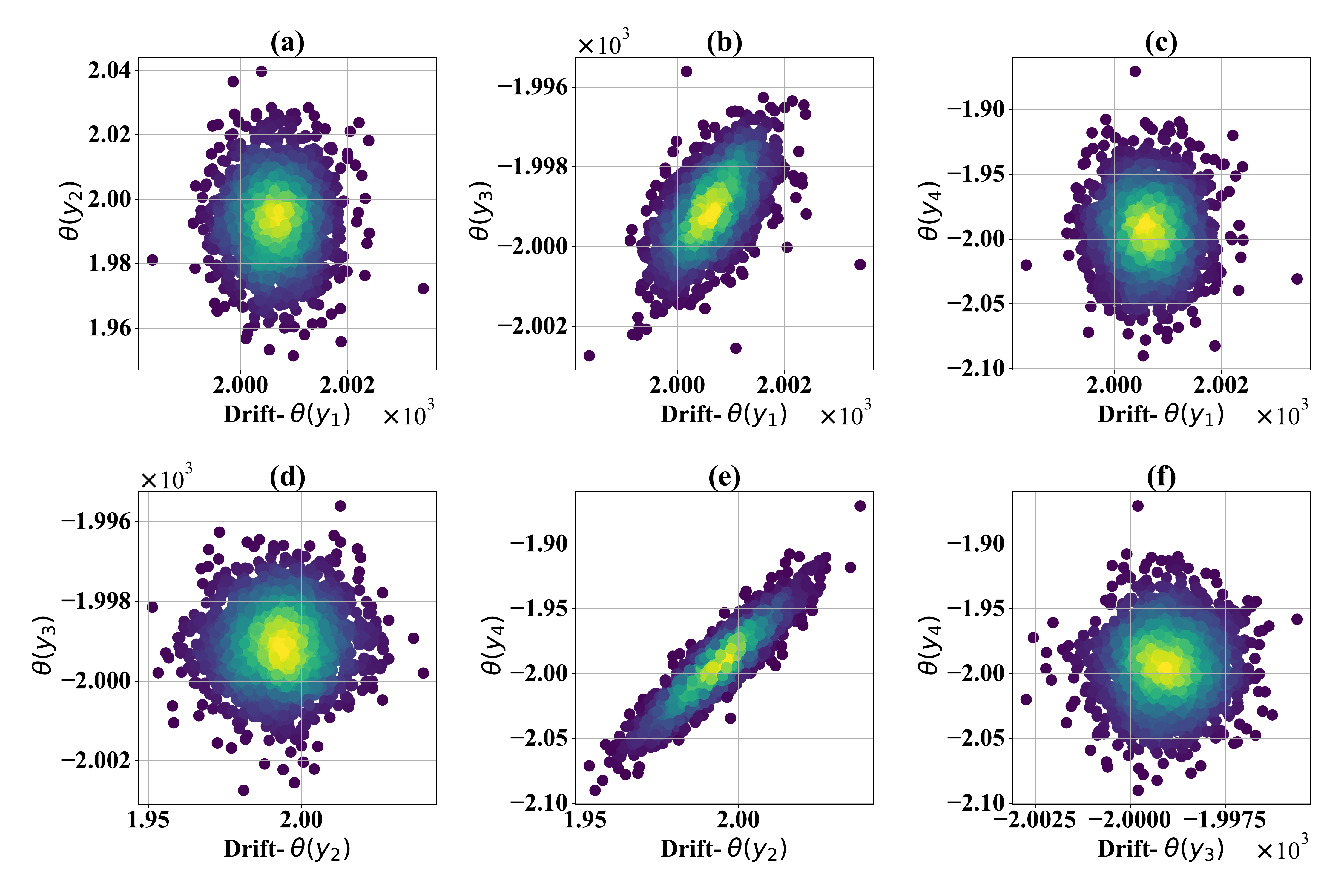}
	\caption{\textbf{Non-linear two-DOF oscillator (second drift component): Pairwise joint posterior distributions of the parameters $\theta_k$ of the selected basis functions}. (a) Joint posterior of the weights $\theta({Y_1})$ and $\theta(Y_2)$ corresponding to the basis functions $\upsilon_1$ and $\upsilon_2$. The yellow color denotes the mean value of the joint distribution. The expected values of the parameters are identified as ${\rm{E}}(\theta (Y_1))$= 2000.65 and ${\rm{E}}(\theta (Y_2))$= 1.99. (b) Joint posterior of the weights $\theta_{Y_1}$ and $\theta_{Y_3}$ corresponding to the basis functions $\upsilon_1$ and $\upsilon_3$. The expected values of the parameters are identified as ${\rm{E}}(\theta (Y_1))$= 2000.65 and ${\rm{E}}(\theta (Y_3))$= -1999.06. (c) Joint posterior of the weights $\theta_{Y_1}$ and $\theta_{Y_4}$ corresponding to the basis functions $\upsilon_1$ and $\upsilon_4$. The expected values of the parameters are identified as ${\rm{E}}(\theta (Y_1))$= 2000.65 and ${\rm{E}}(\theta (Y_4))$= -1.99. (d) Joint posterior of the weights $\theta_{Y_2}$ and $\theta_{Y_3}$ corresponding to the basis functions $\upsilon_2$ and $\upsilon_3$. The expected values of the parameters are identified as ${\rm{E}}(\theta (Y_2))$= 1.99 and ${\rm{E}}(\theta (Y_3))$= -1999.06. (e) Joint posterior of the weights $\theta_{Y_2}$ and $\theta_{Y_4}$ corresponding to the basis functions $\upsilon_2$ and $\upsilon_4$. The expected values of the parameters are identified as ${\rm{E}}(\theta (Y_2))$= 1.99 and ${\rm{E}}(\theta (Y_4))$= -1.99. (f) Joint posterior of the weights $\theta_{Y_3}$ and $\theta_{Y_4}$ corresponding to the basis functions $\upsilon_3$ and $\upsilon_4$. The expected values of the parameters are identified as ${\rm{E}}(\theta (Y_3))$= -1999.06 and ${\rm{E}}(\theta (Y_4))$= -1.99. The weights $\{\theta_{Y_1}, \theta_{Y_2}, \theta_{Y_3}, \theta_{Y_4}\}$ represents the relative value of the parameters $\{k_1, k_2, k_3, k_4\}$. The negative values arises due to the coupling of the systems states. }
	\label{fig2dof_param_drift2}
\end{figure}

The responses of civil engineering structures often exhibit the characteristics of stochastic non-linear dynamical systems due to the external excitation. Here, a 2 story base isolated structure is considered. The system is a linear base isolated structure where the structure is connected to the foundation through a Coulomb friction-type base isolator. Under this considerations, the governing equation of motion of the system is written as,
\begin{equation}
    \begin{array}{l}
{m_1}{{\ddot X}_1}(t) + {c_1}{{\dot X}_1}(t) + \mu {m_1}g{\mathop{\rm sgn}} \left({{\dot X}_1}(t)\right) + {k_1}{X_1}(t) + {c_2}\left( {{{\dot X}_1}(t) - {{\dot X}_2}(t)} \right) + {k_2}\left( {{X_1}(t) - {X_2}(t)} \right) = {\sigma _1}{{\dot B}_1}(t) \\
{m_2}{{\ddot X}_2}(t) + {c_2}\left( {{{\dot X}_2}(t) - {{\dot X}_1}(t)} \right) + {k_2}\left( {{X_2}(t) - {X_1}(t)} \right) = {\sigma _2}{{\dot B}_2}(t)\\
{\bm X}(t=t_0)={\bm X}_0; \quad t \in [0,T],
\end{array}
\end{equation}
where $\mu {m_1}g{\mathop{\rm sgn}} ({{\dot X}_1}(t))$ is the Coulomb friction force arising due to the sliding of bearings in Coulomb oscillator, and $\{m_i, i=1,2\}$, $\{c_i, i=1,2\}$, and $\{k_i, i=1,2\}$ are the mass, damping, and stiffness of $i^{th}$-DOF. Here, $\sigma_1 \ge 0$ and $\sigma_2 \ge 0$ are the strength of the white noise ${\dot B}_1(t)$ and ${\dot B}_2(t)$, where $B_1(t)$ and $B_2(t)$ are the independent Brownian motion. In this case study, the equation discovery involves identification of two drift and two covariance terms (one for each of the DOFs). Following to the case study of the Duffing-Van der pol oscillator, the target linear variation vector for discovery of $i^{th}$; $\forall i = 1, \ldots, 2$ drift term needs to be constructed from the velocity component of the corresponding DOF. For discovery of the $(ij)^{th}$; $\forall i,j = 1, \ldots, 2 $ component of the covariance matrix, the quadratic variation matrix is constructed from the pointwise product of $i^{th}$ and $j^{th}$ response vectors.
\begin{equation}\label{cov_2dof}
    \Gamma  = \left[ {\begin{array}{*{20}{c}}
0&0&0&0\\
0&{{\sigma _1^2}/{{m_1^2}} }&0&0\\
0&0&0&0\\
0&0&0&{\sigma _2^2}/{{m_2^2}}
\end{array}} \right].
\end{equation}
The results for the identification of corresponding basis functions for both the drift terms are presented in Figs. \ref{figdrift}(d) and (e). From Fig. \ref{figdrift}(d), it can be observed that the identified basis functions are $\left\{Y_1(t), Y_2(t), Y_3(t), Y_4(t), sgn(Y_2(t)) \right\}$, which correctly matches with the terms in the equation of motion of first DOF. Similarly, from Fig. \ref{figdrift}(e) it can be verified that the proposed scheme has correctly identified the basis functions of the second drift component as $\left\{Y_1(t), Y_2(t), Y_3(t), Y_4(t) \right\}$. Figs. \ref{fig2dof_param_drift1} and \ref{fig2dof_param_drift2} show the pairwise joint posterior distributions of the weights associated with identified first and second drift basis functions, respectively. From these figures it is evident that the proposed scheme is able to identify the parameters of the basis functions as listed in Table \ref{table_param} with sufficient accuracy. In order to find the explicit values of system stiffness and damping parameters, the parameters of second DOF can be identified from the Figs. \ref{fig2dof_param_drift1}(h), \ref{fig2dof_param_drift1}(e), and \ref{fig2dof_param_drift1}(f).

\begin{figure}[htbp!]
	\centering
	\includegraphics[width=\textwidth]{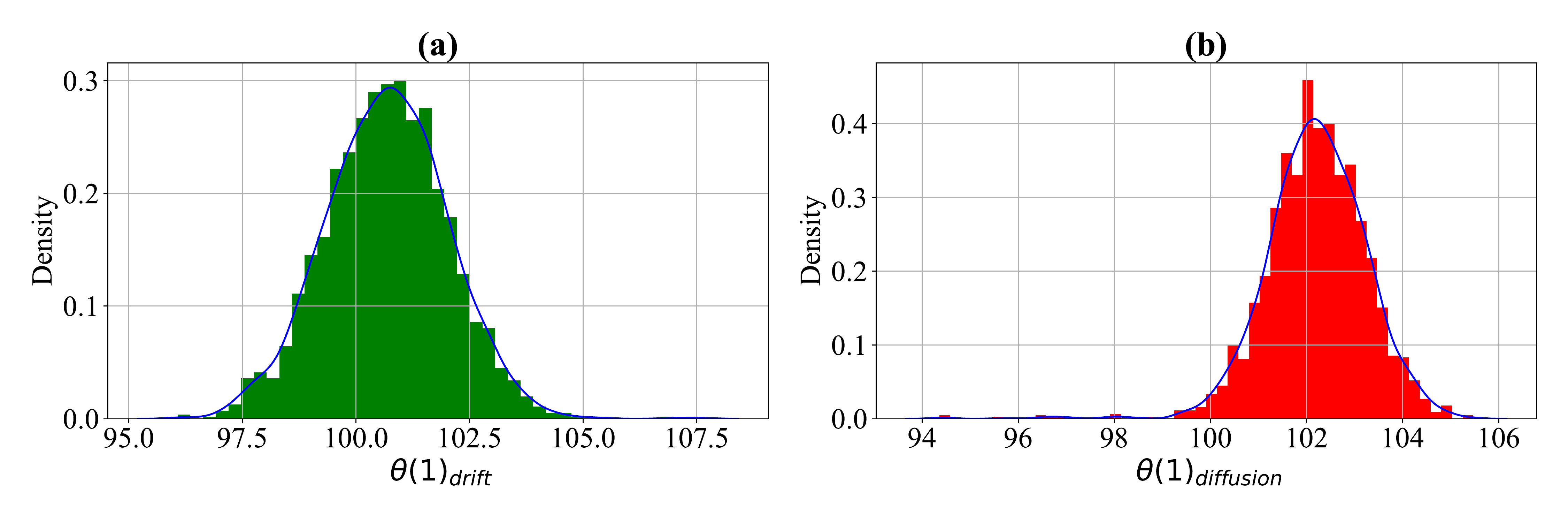}
	\caption{\textbf{Two-DOF shear building: posterior distributions of the weights associated with the discovered basis functions of diffusion components}. (a) First DOF: posterior distribution of the basis function $\theta_1$. The mean of the distribution denotes the expected value of $\sigma_1^2/m_1^2$, which in this case is observed from the figure as approximately 101. (b) Second-DOF: posterior distribution of the basis function $\theta_1$. The mean is observed as 102, which is equivalent to the term $\sigma_2^2/m_2^2$. By performing the square root operation over the mean values one can easily obtain the desired diffusion components.}
	\label{fig2dof_param_diff}
\end{figure}

The identification results for the basis functions along with the posterior distribution of their parameters for both the diffusion terms are depicted in Fig. \ref{fig2dof_param_diff}. Figs. \ref{fig2dof_param_diff}(a) and \ref{fig2dof_param_diff}(b) present the results of the first diffusion term. The identification results of the second diffusion term are shown in Figs. \ref{fig2dof_param_diff}(c) and \ref{fig2dof_param_diff}(d). In both the cases, it is evident that the proposed scheme has correctly identified the basis function as $1$. The posterior distributions of the covariance terms, ${\bf{\Gamma}}^{11}$ and ${\bf{\Gamma}}^{22}$ is plotted in Figs. \ref{fig2dof_param_diff}(b) and \ref{fig2dof_param_diff}(d) whose mean values represents ${{\sigma _1^2}}/{{m_1^2}}$ and ${{\sigma _2^2}}/{{m_2^2}}$. The summery of the results are presented in Table \ref{table_summery}.

\subsection{Case study: Non-linear system with partially observed state variables}
\begin{figure}[htbp!]
	\centering
	\includegraphics[width=\textwidth]{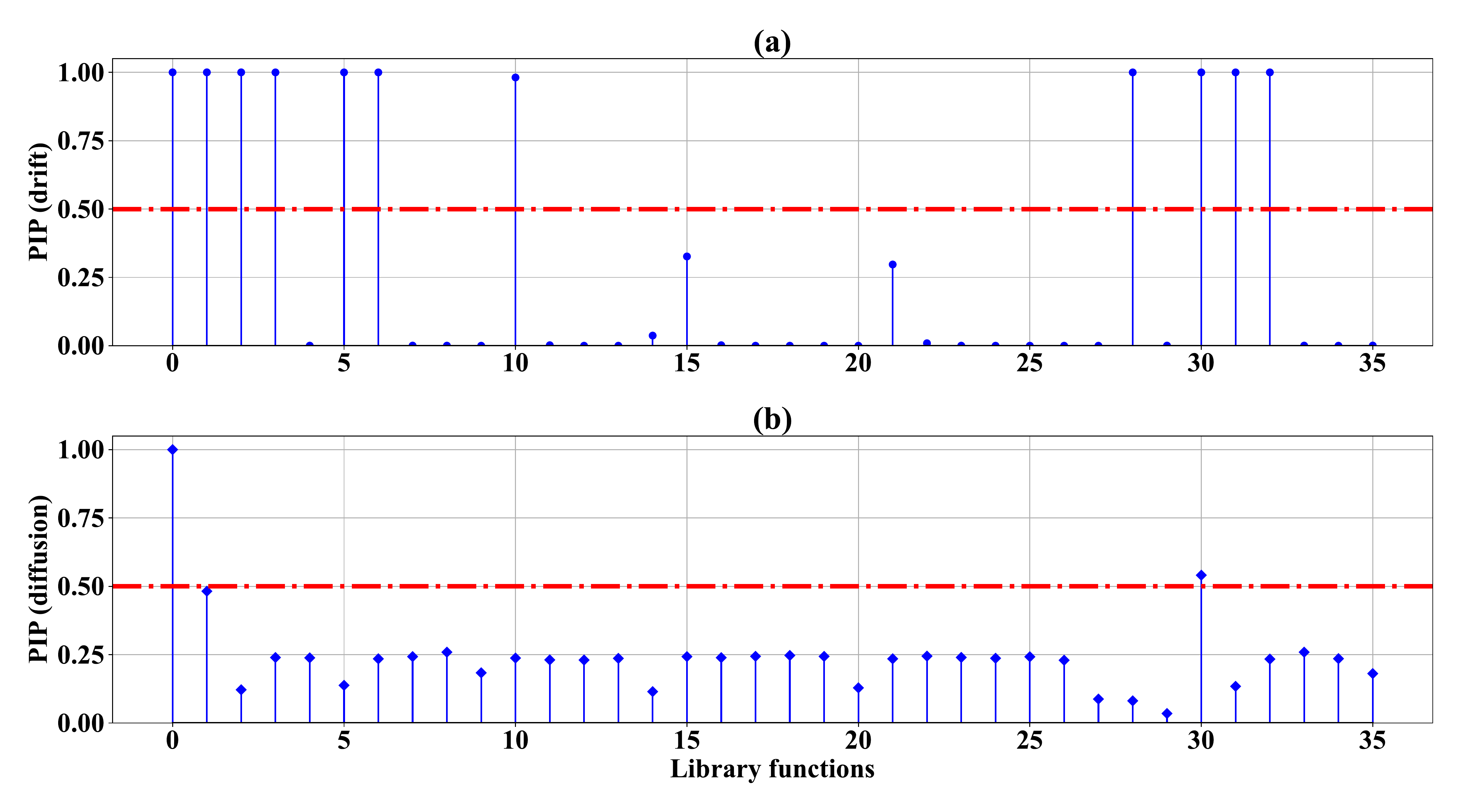}
	\caption{\textbf{Bouc-Wen oscillator: Basis function selection for the Bouc-Wen base isolator system based on marginal posterior inclusion probability (PIP), P$(Z_{k}=1|\bm{Y}); k=1,\ldots,36$}. (a) The identified basis functions for the drift component. The black horizontal axes represent the collection of 36 basis functions, and the red line represents marginal PIP $>$ 0.5. The identified basis functions are $\{\upsilon(0), \upsilon(1), \upsilon(2), \upsilon(3), \upsilon(5), \upsilon(6), \upsilon(10), \upsilon(28), \upsilon(30), \upsilon(31), \upsilon(32) \}$. These basses corresponds to the functions $\{ 1, {X_1}, {X_2}, {X_1^2}, {X_2^2}, {X_1^3}, {X_1^4}, {sgn({X_1})}, {\left| {{X_1}} \right|}, {\left| {{X_2}} \right|}, {{X_1}\left| {{X_1}} \right|} \}$. (b) The identified basis functions for the diffusion component. The identified function are $\{1, {X_1}, {\left| {{X_1}} \right|} \}$.}
	\label{figboucwen_basis}
\end{figure}

In this section, a case study has been undertaken where we do not have access to all the state variables. We consider a Bouc-Wen oscillator where the system is described using three states namely, displacement and velocity of the main mass, and the hysteresis displacement associated with the isolator \cite{tripura2020ito,tripura2020real}. In this context, we note that the measurement of the hysteresis displacement is practically intractable and only the displacement and velocity states of the main system is available. Further, measuring the velocity  of a mechanical system using sensors is a practically challenging task. Through this example, we illustrate the applicability of the proposed approach for such a system. The governing equation of motion of a sdof Bouc-Wen oscillator is \cite{tripura2020ito}:
\begin{equation}
    m\ddot X(t) + c\dot X(t) + k\lambda X(t) + k\left( {1 - \lambda } \right)Z(t) = \sigma \dot B(t)
    \qquad
    {\bm X}(t=t_0)={\bm X}_0; \quad t \in [0,T],
\end{equation}
where $m$, $c$, and $k$ are the mass, damping, and stiffness of the system, $X(t)$ is the system state, $\lambda$ is a factor that defines the participation of the elastic force $F_e$ and hysteresis force $F_h$, $F_e(t) = k\lambda X(t)$, $F_h(t) = k\left( {1 - \lambda } \right)Z(t)$, and $Z(t)$ is the hysteresis displacement. The evolution of the hysteresis parameter is defined using the non-linear differential equation:
\begin{equation}
    Z(t) =  - {A_1}Z(t)\left| {\dot X(t)} \right|{\left| {Z(t)} \right|^{\bar n - 1}} - {A_2}\dot X(t){\left| {Z(t)} \right|^{\bar n}} + {A_3}\dot X(t)
    \qquad
    {\bm Z}(t=t_0)={\bm Z}_0; \quad t \in [0,T],
\end{equation}
The positive exponential parameter $n$ defines the smoothness of the transition from elastic to the post-elastic branch, and the parameters $A_1$, $A_2$, and $A_3$ control the size and shape of the hysteresis loop. 
The drift and diffusion terms are:
\begin{equation}
    f(t,{{\bm{X}}_t}) = \left[ {\begin{array}{*{20}{c}}
{{X_2}(t)}\\
{\dfrac{{ - 1}}{m}\left( {k\lambda {X_1}(t) + c{X_2}(t) + k(1 - \lambda ){X_3}(t)} \right)}\\
{ - {A_1}{X_3}(t)\left| {{X_2}(t)} \right|{{\left| {{X_3}(t)} \right|}^{\bar n - 1}} - {A_2}{X_2}(t){{\left| {{X_3}(t)} \right|}^{\bar n}} + {A_3}{X_2}(t)}
\end{array}} \right]; \quad g(t,{{\bm{X}}_t}) = \left[ {\begin{array}{*{20}{c}}
0\\
{\dfrac{\sigma }{m}}\\
0
\end{array}} \right].
\end{equation}
Then, the covariance of the diffusion matrix is obtained as,
\begin{equation}
    \Gamma  = g(t,{{\bf{X}}_t})g{(t,{{\bf{X}}_t})^T} = \left[ {\begin{array}{*{20}{c}}
0&0&0\\
0&{\dfrac{{{\sigma ^2}}}{{{m^2}}}}&0\\
0&0&0
\end{array}} \right].
\end{equation}
For the identification of the system only the displacement value (noisy) is considered as input to the algorithm.
The system is simulated for $T$ = 1s at a frequency of 1000Hz using the parameter values, $m$ = 1, $c$ = 20, $k$ = 100000, $\lambda$ = 0.5, $A_1$ = 0.5, $A_2$ = 0.5, $A_3$ = 1, $\bar{n} = 3$, and $\sigma_1$ = 2. The identified basis functions for the drift and diffusion terms are presented in Fig. \ref{figboucwen_basis}. In sub-figure (a) it is evident that there are more number of basis functions than the input equation. It is straightforward to note that this extra basis functions arises to take care of the non-observable hysteresis parameter $Z(t)$. One can then identify the basis functions ${\Theta _f}({\bf{X}})$ and ${\Theta _g}({\bf{X}})$ for the drift and diffusion, respectively, that best represent the data as,
\begin{equation}
\begin{array}{l}
{\Theta _f}({\bf{X}}) = \left[ {\begin{array}{*{20}{c}}
1&{{X_1}}&{{X_2}}&{X_1^2}&{X_2^2}&{X_1^3}&{X_1^4}&{sgn({X_1})}&{\left| {{X_1}} \right|}&{\left| {{X_2}} \right|}&{{X_1}\left| {{X_1}} \right|}
\end{array}} \right]\\
{\Theta _g}({\bf{X}}) = \left[ {\begin{array}{*{20}{c}}
1&{{X_1}}&{\left| {{X_1}} \right|}
\end{array}} \right].
\end{array}
\end{equation}

\begin{figure}[t!]
	\centering
	\includegraphics[width=\textwidth]{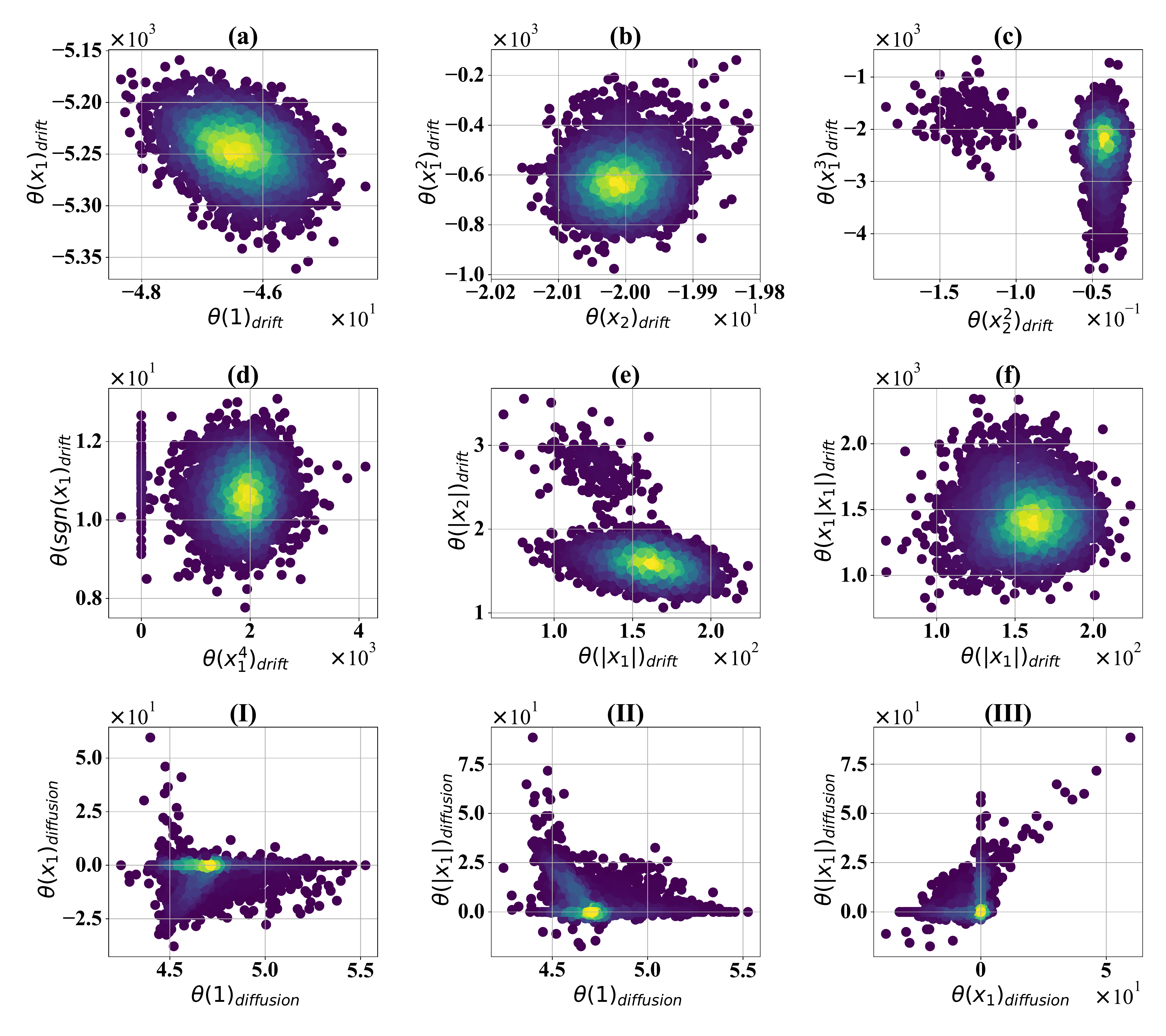}
	\caption{\textbf{Bouc-Wen oscillator: Pairwise joint posterior distributions of the parameters $\theta_k$ of the selected basis functions}. The subplots (a)$\to$(f) presents the joint posterior distributions of drift parameters whereas the same is presented in the subplots (I)$\to$(III). (a) Joint posterior distribution of $\theta (1)$ and $\theta (X_1)$. The parameters are identified as ${\rm{E}}(\theta (1))$= -46.4 and ${\rm{E}}(\theta (X_1))$= -5246. (b) Joint posterior distribution of $\theta (X_2)$ and $\theta (X_1^2)$. The expected value of the parameters are identified as ${\rm{E}}(\theta (X_2))$= -20 and ${\rm{E}}(\theta (X_1^2))$= -614. (c) Joint posterior distribution of $\theta (X_2^2)$ and $\theta (X_1^3)$. The parameters are identified as ${\rm{E}}(\theta (X_2^2))$= -0.05 and ${\rm{E}}(\theta (X_1^3))$= -2328. (d) Joint posterior distribution of $\theta (X_1^4)$ and $\theta (sgn(X_1))$. The expected value of the parameters are identified as ${\rm{E}}(\theta (X_1^4))$= 1829 and ${\rm{E}}(\theta (sgn(X_1)))$= 10.64. (e) Joint posterior distribution of $\theta (|X_1|)$ and $\theta (|X_2|)$. The parameters are identified as ${\rm{E}}(\theta (|X_1|))$= 159 and ${\rm{E}}(\theta (sgn(X_1)))$= 10.64. (f) Joint posterior distribution of $\theta (|X_1|)$ and $\theta (X_1|X_1|)$. The parameters are identified as ${\rm{E}}(\theta (|X_1|))$= 159 and ${\rm{E}}(\theta (X_1|X_1|))$= 1439. (I) The joint posterior of the weights $\theta_{1}$ and $\theta_2$ corresponding to the basis functions $1$ and $X_1$ is shown. The yellow color denotes the mean values of the joint distribution. (II) The joint posterior of the weights $\theta_{1}$ and $\theta_{30}$ corresponding to the basis functions $1$ and $|X_1|$ is shown. (III) The joint posterior of the weights $\theta_{2}$ and $\theta_30$ corresponding to the basis functions $X_1$ and $|X_1|$ is shown.}
	\label{figboucwen_posterior}
\end{figure}
The joint posterior distributions of the retained basis functions in Fig. \ref{figboucwen_basis} (a) and (b) are plotted in the Fig. \ref{figboucwen_posterior}. Based on the basis selection in Fig. \ref{figboucwen_basis} and from the joint posteriors of the parameter $\theta_i$ plotted in Fig. \ref{figboucwen_posterior}, the final drift and diffusion fields for the Bouc-Wen system can identified as follows,
\begin{equation}
    \begin{array}{ll}
f(t,X) =&  - 46.4 - 5246{X_1} - 20{X_2} - 614X_1^2 - 0.05X_2^2 - 2328X_1^3 + 1829X_1^4 + 10.64sgn({X_1}) + 159\left| {{X_1}} \right| \\
& + 1.625\left| {{X_2}} \right| + 1439{X_1}\left| {{X_1}} \right|\\
g(t,X) =& 4.65 - 4.872{X_1} + 6.786\left| {{X_1}} \right|.
\end{array}
\end{equation}

\begin{table}[ht!]
    \centering
    \begin{tabular}{ll}
        \hline
        \textbf{Correct SDEs}: & $\left\{ \begin{array}{l}
                d{X_2}(t) =  - \left( {5000{X_1}(t) + 20{X_2}(t) + 5000{X_3}(t)} \right)dt + 2dB(t)\\
                d{X_3}(t) = \left( { - 0.5{X_3}(t)\left| {{X_2}(t)} \right|{{\left| {{X_3}(t)} \right|}^2} - 0.5{X_2}(t){{\left| {{X_3}(t)} \right|}^3} + {X_2}(t)} \right)dt
                \end{array} \right.$\\
        \hline 
        \textbf{Identified SDEs}: & $\left\{ \begin{array}{ll}
                d{X_2}(t) =& \Bigl(  - 46.4 - 5246{X_1} - 20{X_2} - 614X_1^2 - 2328X_1^3 + 1829X_1^4 + 10.64sgn({X_1}) + 159\left| {{X_1}} \right| \\& + 1.63\left| {{X_2}} \right| + 1439{X_1}\left| {{X_1}} \right|  \Bigr)dt + \sqrt {4.65 - 3.79{X_1} + 5.76\left| {{X_1}} \right|} dB(t) \end{array} \right.$\\
        \hline
    \end{tabular}
    \caption{\textbf{Identification results of the Bayesian physics discovery for Bouc-Wen oscillator}}
    \label{table_boucwen}
\end{table}

\begin{figure}[htbp!]
	\centering
	\includegraphics[width=\textwidth]{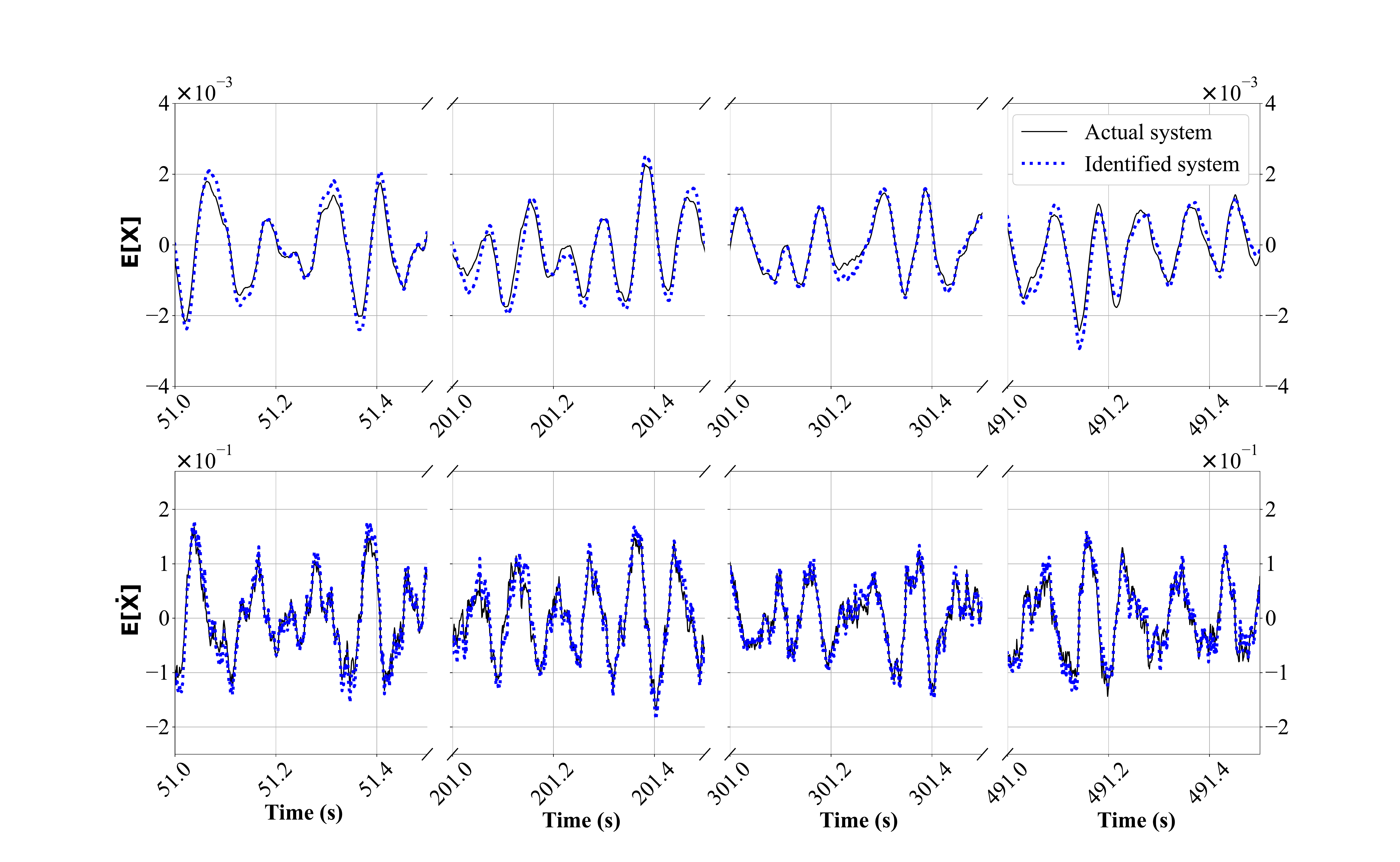}
	\caption{\textbf{Bouc-Wen oscillator: Future prediction in an unseen scenario using the discovered equation in the absence of the hysteresis data}. The governing equation of the Bouc-Wen system learned using 1 second of data sampled at 1000Hz. The prediction is performed for 500 seconds ahead of time using the discovered Bouc-Wen model. In the figure, the ensemble of the displacement and velocity prediction data over 500 seconds is presented. The simulation is performed using a different Brownian noise intensity of $\sigma$ = 6, which is significantly higher than the training data. The dotted red line shows the prediction using the discovered equation and the solid blue line denotes the accurate simulated data. The predicted data almost accurately emulates the original system even though the measurements for all the states are not available. }
	\label{figboucwen_time}
\end{figure}

For further treatment the term $0.05X_2^2$ is neglected due to its non-significant parameter value. With this, the identified equation for the Bouc-Wen system is shown in Eq. \eqref{identboucwen}:
\begin{multline}\label{identboucwen}
    \ddot X + 20\dot X + 5246X + 614{X^2} + 2328{X^3} - 1829{X^4} - 10.64sgn(X) - 159\left| X \right| \\
    - 1.625\left| {\dot X} \right| - 1439X\left| X \right| + 46.4 = 4.65 - 4.872X + 6.786\left| X \right|.
\end{multline}
In order to verify the accuracy of the identified system, the identified equation (Eq. \eqref{identboucwen}) is simulated using the EM scheme with a time step of $\Delta t$ = 0.001. To account for the fact that in future the external excitation will be uncertain and the intensity of the same will not remain same, we model the forcing function using Brownian force with different intensity. To be specific, a Brownian noise intensity of $\sigma$ = 6 is used to simulate the new unseen environment. The results are compared in Fig. \ref{figboucwen_time}.
We observe that the results obtained using the identified physics matches almost exactly with the ground truth obtained by solving the actual system. It is worthwhile to note that the proposed approach require only 1s of data and is able to predict the the responses for a stochastic force of different intensity at distant future (500s). This clearly indicates that the proposed approach, even with partially observed displacement only measurement, generalizes to unseen environment and provides accurate estimate even for out-of-distribution inputs. 
The summary of the cases studies undertaken in this work is presented in the Table \ref{table_summery}. 

\section{Discussions}\label{sec:conclusion}
We proposed a novel data-driven framework for discovering governing physics of multi-dimensional non-linear dynamical systems from limited and noisy output only data. In many natural process the estimate of external input is intractable due to the limitations in the present measurement technologies. Further, measurement of all the state variables is often not possible due to high operational cost; the proposed framework is developed to identify governing equations in such scenarios. In essence, the proposed approach identifies SDEs from displacement only measurements; in other words, no measurement for the input is required. The deterministic drift component of the SDEs captures the dynamics of the underlying dynamical system whereas the external stochastic force is identified in terms of the diffusion component. This is a key novelty of the proposed framework. 

The proposed framework employs the sparse Bayesian linear regression in conjunction with the Gibbs sampler to simultaneously obtain the basis functions of the model and their associated parameters. To that end, the Kramers-Moyal expansion is utilized to express the drift and diffusion dynamics of an SDE in terms of the measured samples paths. The drift and diffusion vectors obtained from the sample paths are then used as a target vector in the sparse regression. Use of the Kramers-Moyal expansion within the purview of sparse Bayesian linear regression for discovery of an explainable governing physics is another novelty of the proposed framework. The unified framework possesses the ability of the Bayesian inference to update the probability of observing the correct basis function based on the new information. The ability to update the information is missing in recently published least-square based physics discovery schemes. In a noisy and unseen environment, one would be more interested in dealing with the probability distribution of a random variable rather than a point estimate; the proposed framework thus, takes a leap in such situations and provides a probability distribution for each of the weights associated with corresponding basis functions. The obtained standard deviations of the distributions can then be used for defining the lower and uppers bounds on the estimates of an unknown parameters. Another aspect of the proposed framework is its advantages over the Neural network based grey box models, where the explicit expression of the discovered governing physics is not known. In cases of partially observed processes, the proposed framework seems to provide an explainable equation for the governing physics. Although the basis functions in the obtained equation is significantly different from the one determined from first principal laws, the obtained model is able to predict the distant future without significant error. To the knowledge of authors, the discovery of physics in a partially observed process from the purview of probability theory is one of the key contributions of the proposed novel framework. 

Despite of the advantages of the proposed framework over presently available physics discovery techniques there are certain issues that requires special attention for accurate identification of the governing physics. The first issue is the judicious selection of the candidate basis function. The presence of basis functions with high correlation may sometimes yield a physical law different from the actual physics. This is evident from the identification example on the Black-Scholes equation. The second issue is the associated computational cost. Although the computational cost of the proposed scheme is significantly less than the time required to train its neural network alternatives, further improvements can be made. One such alternative is to use the marginal standard deviation criteria, instead of taking the mean of the latent vector. Another improvement in terms of computational efficiency can be made by devising an appropriate filter to process out the signal noise synchronously with the sparse Bayesian linear regression. The third issue is related to the quality of the data. The low-fidelity data are less expensive to obtain but the representation of the physical model can be poor due to the presence of noise. On the other hand, the  high-fidelity data are highly accurate but expensive to obtain. The proposed framework in its present state is not well equipped to make use of both the low and high-fidelity data. Thus, to further enhance the fidelity of the proposed framework for field applications a more robust algorithm needs to be formulated. The use of low-fidelity data ensure low operational cost while the high-fidelity data will ensure the accuracy of the discovered physics. Lastly, the proposed framework leverages the Kramers-Moyal expansion to express the drift and diffusion components of an SDE in terms of the first and second order moments of the sample paths. The moments are approximated in a limiting sense by making use of the absolute and quadratic identities of Brownian motion. In this context, future improvements in the proposed framework can be made by exploiting the higher order moments to retain the higher order terms from Stochastic-Taylor expansions of the drift and diffusion terms. Overall, the proposed framework provides a novel methodology for discovering the governing physics from displacement only measurements and is particularly useful whenever it is not possible to measure the input force. The learned equations have shown to be accurate  and demonstrated good prediction ability in distant future.

\appendix

\appendix

\section{Kramers-Moyal expansion for estimation of the drift and diffusion terms of an SDE from sample paths}\label{appenA}
Let us consider, $p(X, t)$=${\rm{P}}(X,t|X_0,t_0)$ to be the transition probability density of the solution of the SDE in Eq. \eqref{sdeg}. Then the Kramers-Moyal expansion is written as \cite{risken1996fokker},
\begin{equation}\label{A1}
    \dfrac{{\partial P(X,t)}}{{\partial t}} = {\sum\limits_{n = 1}^\infty  {\left( { - \dfrac{\partial }{{\partial X}}} \right)} ^n}{D^{(n)}}(X,t)p(X,t),
\end{equation}
where the coefficients in the expansion are given as,
\begin{equation}
    {D^{(n)}}(X) = {\left. {{{\left. {\dfrac{1}{{n!}}{M_n}(t)} \right|}_{t = 0}} = \dfrac{1}{{n!}}\mathop {\lim }\limits_{{\Delta t}  \to 0} \dfrac{1}{{\Delta t} }\left\langle {{{\left| {X(t + {\Delta t} ) - z} \right|}^n}} \right\rangle } \right|_{X(t) = z}}.
\end{equation}
In order to know how many terms in Eq. \eqref{A1} will be active for the SDE in Eq. \eqref{sdeg}, a Fokker-Planck equation for the pdf of the solution of Eq. \eqref{sdeg} needs to be constructed. Towards this, let us consider a well behaved generic function $F(X,t)$ which is at least twice differential. For this function, the It\^{o}'s lemma is written as,
\begin{equation}
    dF(X,t) = \dfrac{{\partial F(X,t)}}{{\partial X}}dX(t) + \dfrac{1}{2}\dfrac{{{\partial ^2}F(X,t)}}{{\partial {X^2}}}{g^2}(X,t)dt.
\end{equation}
Substituting Eq. \eqref{sdeg} in the above equation gives the following:
\begin{equation}\label{itolemma2}
    dF(X,t) = \left( {\dfrac{{\partial F(X,t)}}{{\partial X}}f(X,t) + \dfrac{1}{2}\dfrac{{{\partial ^2}F(X,t)}}{{\partial {X^2}}}{g^2}(X,t)} \right)dt + \dfrac{{\partial F(X,t)}}{{\partial X}}g(X,t)dB(t).
\end{equation}
Upon taking the derivative with respect to $t$ on both sides yields:
\begin{equation}
    \dfrac{{dE[F(X,t)]}}{{dt}} = E\left( {\dfrac{{\partial F(X,t)}}{{\partial X}}f(X,t) + \dfrac{1}{2}\dfrac{{{\partial ^2}F(X,t)}}{{\partial {X^2}}}{g^2}(X,t)} \right).
\end{equation}
Expanding the above equation using expectation operator and noting that the last term in the expression is a stochastic integral, one can invoke the result ${\rm{E}}\left[ {\int_0^t {\dfrac{{\partial F(X,s)}}{{\partial X}}g(X,s)dB(s)} } \right] = 0$, where ${\rm{E}}(\cdot)$ is the expectation operator to obtain the following,
\begin{equation}
    \int\limits_\infty ^\infty  {p(X,t)\dfrac{{dF(X)}}{{dt}}dX}  = \int\limits_\infty ^\infty  {p(X,t)\left( {\dfrac{{\partial F(X,t)}}{{\partial X}}f(X,t) + \dfrac{1}{2}\dfrac{{{\partial ^2}F(X,t)}}{{\partial {X^2}}}{g^2}(X,t)} \right)dX}. 
\end{equation}
By expanding the terms in the above equation using integration by parts and assuming that $\mathop {\lim }\limits_{X \to 0} p(X,t) \to 0$, the following weak form is obtained:
\begin{equation}
    \int\limits_\infty ^\infty  {F(X)\left( {\dfrac{{\partial p(X,t)}}{{\partial t}} + \dfrac{{\partial (p(X,t)f(X,t))}}{{\partial X}} - \dfrac{1}{2}\dfrac{{{\partial ^2}(p(X,t){g^2}(X,t))}}{{\partial {X^2}}}} \right)dX}  = 0.
\end{equation}
As the function $F(X,t)$ is arbitrarily selected, the Fokker-Plank-Kolmogorov equation is obtained.
\begin{equation}
    \dfrac{{\partial p(X,t)}}{{\partial t}} =  - \dfrac{{\partial (p(X,t)f(X,t))}}{{\partial X}} + \dfrac{1}{2}\dfrac{{{\partial ^2}(p(X,t){g^2}(X,t))}}{{\partial {X^2}}};\;p(0,X) = {p_0}(X).
\end{equation}
In a general setting the above equation for higher dimensional diffusion process is expressed as,
\begin{equation}
    \dfrac{{\partial p(X,t)}}{{\partial t}} = {L_t}p(X,t) =  - \sum\limits_i^m {\dfrac{{\partial (p(X,t){f_i}(X,t))}}{{\partial {X_i}}}}  + \dfrac{1}{2}\sum\limits_i^m {\sum\limits_j^m {\sum\limits_k^n {\dfrac{{{\partial ^2}(p(X,t){g_{i,k}}(X,t){g_{j,k}}(X,t))}}{{\partial {X_i}\partial {X_j}}}} } },
\end{equation}
where the random variable $X$ satisfies the SDE in Eq. \eqref{sdeg}. At this point it is clear that there will be two terms active in  Eq. \eqref{sdeg}. Thus for $n$=2, Eq. \eqref{A1} yields,
\begin{equation}\label{A10}
    \dfrac{{\partial P(X,t)}}{{\partial t}} =  - \dfrac{\partial }{{\partial X}}\left[ {{D^{(1)}}(X,t)p(X,t)} \right] + \dfrac{{{\partial ^2}}}{{\partial {X^2}}}\left[ {{D^{(2)}}(X,t)p(X,t)} \right].
\end{equation}
On comparison of Eqs. \eqref{A1} and \eqref{A10}, it is straightforward to note that to estimate the drift and diffusion terms it suffices to estimate the first and second order moments of the variations of the random variable $X(t)$ and then calculate the coefficients ${D^{(1)}}$ and ${D^{(2)}}$, formally,
\begin{equation}
    f(X,t) = {D^{(1)}},\; g^2(X,t) = {D^{(2)}}.
\end{equation}
To derive these coefficients in the Kramers-Moyal expansion, it is imperative to understand the one-step It\^{o}-Taylor expansion of the random variable $X(t)$. Referring to the Eq. \eqref{itolemma2}, the integral form the It\^{o}-lemma can be expressed as follows,
\begin{equation}\label{ito_intgral}
    F({X_{t + h}},t + h) = F({X_t},t) + \int_t^{t + h} {\left\{ {f({X_s},s)F'({X_s},s) + \dfrac{1}{2}{g^2}({X_s},s)F''({X_s},s)} \right\}ds + } \int_t^{t + h} {g({X_s},s)F'({X_s},s)dB(s)}, 
\end{equation}
where $F'({X_s},s)$ and $F''({X_s},s)$ denotes the first and second order partial derivatives with respect to system states. For simplicity, two stochastic operators are defined in the following form:
\begin{equation}\label{operator}
    \begin{array}{l}
{\Im ^0}(.) = \dfrac{{\partial (.)}}{{\partial t}} + \sum\limits_i^m {{f_i}({X_t},t)\dfrac{{\partial (.)}}{{\partial {X_i}}}}  + \dfrac{1}{2}\sum\limits_i^m {\sum\limits_j^m {\sum\limits_k^n {{g_{i,k}}({X_t},t){g_{j,k}}({X_t},t)\dfrac{{{\partial ^2}(.)}}{{\partial {X_i}\partial {X_j}}}} } } \\
{\Im ^1}(.) = \sum\limits_i^m {\sum\limits_k^n {{g_{i,k}}({X_t},t)\dfrac{{\partial (.)}}{{\partial {X_i}}}} }. 
\end{array}
\end{equation}
Using the operators, Eq. \eqref{ito_intgral} can be rephrased as,
\begin{equation}
    F({X_{t + h}},t + h) = F({X_t},t) + \int_t^{t + h} {{\Im ^0}\left( {F({X_s},s)} \right)ds + } \int_t^{t + h} {{\Im ^1}\left( {F({X_s},s)} \right)dB(s)} .
\end{equation}
Then substituting $F({X_t},t) = X(t)$, one verifies that ${\Im ^0}X(t) = f({X_t},t)$, and ${\Im ^1}X(t) = g({X_t},t)$. This yields the first iteration:
\begin{equation}\label{iteration1}
    \begin{array}{ll}
     X(t + h) &= X(t) + \int_t^{t + h} {{\Im ^0}\left( {X(s)} \right)ds + } \int_t^{t + h} {{\Im ^1}\left( {X(s)} \right)dB(s)} \\
     &= X(t) + \int_t^{t + h} {f({X_s},s)ds + } \int_t^{t + h} {g({X_s},s)dB(s)}.
    \end{array} 
\end{equation}
In order to perform the second iteration it is required to find the stochastic expansion of the terms $F({X_t},t) = f({X_s},s)$, and $F({X_t},t) = g({X_s},s)$. Thus, expanding $f({X_t},t)$, and $g({X_t},t)$ and using the operators in Eq. \eqref{operator} yields,
\begin{equation}
\begin{array}{l}
f({X_s},s) = f(X,t) + \int_t^{{s_1}} {{\Im ^0}\left( {f({X_{{s_2}}},{s_2})} \right)d{s_2} + } \int_t^{{s_1}} {{\Im ^1}\left( {f({X_{{s_2}}},{s_2})} \right)dB({s_2})} \\
g({X_s},s) = g(X,t) + \int_t^{{s_1}} {{\Im ^0}\left( {g({X_{{s_2}}},{s_2})} \right)d{s_2} + } \int_t^{{s_1}} {{\Im ^1}\left( {g({X_{{s_2}}},{s_2})} \right)dB({s_2})} .
\end{array}
\end{equation}
On substituting the above result in Eq. \eqref{iteration1} and further iterating,
\begin{equation}\label{final_itotaylor}
X(t + h) - X(t) = f(X,t)\int_t^{t + h} {d{s_1} + } g(X,t)\int_t^{t + h} {dB({s_1}) + } {\Im ^1}\left( {g(X,t)} \right)\int_t^{t + h} {\int_t^{{s_1}} {dB({s_2})dB({s_1})} }  + R,
\end{equation}
where the remainder term $R$ is,
\begin{equation}
    \begin{array}{ll}
R =& \int_t^{t + h} {\int_t^{{s_1}} {{\Im ^0}\left( {f({X_{{s_2}}},{s_2})} \right)d{s_2}d{s_1} + } } \left\{ {\int_t^{t + h} {\int_t^{{s_1}} {{\Im ^1}\left( {f({X_{{s_2}}},{s_2})} \right)dB({s_2})d{s_1}} }  + \int_t^{t + h} {\int_t^{{s_1}} {{\Im ^0}\left( {g({X_{{s_2}}},{s_2})} \right)d{s_2}dB({s_1})} } } \right\} + \\
& \int_{{t_0}}^t {\int_{{t_0}}^{{s_1}} {\int_t^{{s_2}} {{\Im ^1}{\Im ^0}\left( {g({X_{{s_3}}},{s_3})} \right)d{s_3}} dB({s_2})dB({s_1})} }  + \int_{{t_0}}^t {\int_{{t_0}}^{{s_1}} {\int_t^{{s_2}} {{\Im ^1}{\Im ^1}\left( {g({X_{{s_3}}},{s_3})} \right)dB({s_3})} dB({s_2})dB({s_1})} } .
\end{array}
\end{equation}
The above equation is infinitely expandable using the Eq. \eqref{operator}. For further treatment, the first moment is taken on both side. Noting the results $\left\langle {\int_t^{t + h} {dB({s_1})} } \right\rangle  = 0$ and $\left\langle {\int_{{t_0}}^t {\int_{{t_0}}^{{s_1}} {dB({s_2})dB({s_1})} } } \right\rangle  = \left\langle {\int_{{t_0}}^t {B({s_1})dB({s_1})}  - B(t)\int_{{t_0}}^t {dB({s_1})} } \right\rangle = 0$,
\begin{equation}
\begin{array}{ll}
\left\langle {X(t + h) - X(t)} \right\rangle  &= f(X,t)h + g(X,t)\left\langle {\int_t^{t + h} {dB({s_1})} } \right\rangle  + {\Im ^1}\left( {g(X,t)} \right)\left\langle {\int_t^{t + h} {\int_t^{{s_1}} {dB({s_2})dB({s_1})} } } \right\rangle  + \left\langle R \right\rangle \\
 &= f(X,t)h + \left\langle R \right\rangle .
\end{array}
\end{equation}
If the number of iteration is $k$, then the Brownian integrals $\int_t^{t + h} {\int_t^{{s_1}} { \ldots \int_t^{{s_k}} {dB({s_{k + 1}}) \ldots } dB({s_2})dB({s_1})} }$ of multiplicity $k$ have a contribution proportional to $h^{k/2}$, the time integrals $\int_t^{t + h} {\int_t^{{s_1}} { \ldots \int_t^{{s_k}} {d{s_{k + 1}} \ldots } d{s_2}d{s_1}} } $ have a contribution proportional to $h^k$, and the combination  shares a contribution between $h^{k/2}$ and $h^{k}$. Thus it is easy to infer that as $h \to 0$, the higher order terms in the remainder $R$ will vanish. After invoking this facts in the above expression, the first coefficient in Kramers-Moyal expansion is obtained as,
\begin{equation}
{D^{(1)}} = f(X,t) = {\left. {\mathop {\lim }\limits_{\Delta t \to 0} \dfrac{1}{{\Delta t}}\left\langle {{X^{(1)}}(t + \Delta t) - z} \right\rangle } \right|_{X(t) = z}}.
\end{equation}
To derive the second coefficient, it is only required to find the quadratic variation of the increment process of $X(t)$. Upon taking second moment on both sides of Eq. \eqref{final_itotaylor}, the following is obtained,
\begin{equation}
    \begin{array}{ll}
\left\langle {{{\left| {X(t + h) - X(t)} \right|}^2}} \right\rangle  &= f(X,t)h + g(X,t)\Delta B + \left\langle R \right\rangle \\
 &= {f^2}(X,t){h^2} + 2f(X,t)g(X,t)h\Delta B + {g^2}(X,t){\left( {\Delta B} \right)^2} + \left\langle R \right\rangle \\
 &= {g^2}(X,t)h + \left\langle R \right\rangle.
\end{array}
\end{equation}
In the above proof, the It\^{o} identities  are utilized, which states that under the mean square convergence theory the quadratic variation of the time and Brownian increments are given as $d{s_2}d{s_1} = 0$, $d{s_1}dW({s_1}) = 0$, $dW({s_2})d{s_1} = 0$, $dW({s_1})dW({s_2}) = 0$. The deterministic time integrals will vanish automatically since they have finite variance and zero quadratic covariance and the other higher order Brownian integrals will vanish as $h \to 0$. Thus the second coefficient in the Kramers-Moyal expansion is obtained as,
\begin{equation}
{D^{(2)}} = g^2(X,t) = {\left. {\dfrac{1}{2}\mathop {\lim }\limits_{\Delta t \to 0} \dfrac{1}{{\Delta t}}\left\langle {{{\left| {{X^{(1)}}(t + \Delta t) - z} \right|}^2}} \right\rangle } \right|_{X(t) = z}}.
\end{equation}
From the ergodic assumption of the evolution of process $X(t)$, the time average $\left\langle \cdot \right\rangle$ is often replaced by the expectation operator $\rm{E}(\cdot)$. For $m$-variables ${\bm{X}} = \left\{ {{X_1},{X_2}, \ldots {X_m}} \right\}$, the diffusion process has the form,
\[    d{X_i}(t) = {f_i}({{\bm{X}}_t},t) + \sum\limits_j^n {{g_{ij}}({{\bm{X}}_t},t)d{B_j}(t)} ;\;i = 1,2, \ldots m.
\]
The properties of the Brownian motion are: $\left\langle {{B_i}(t)} \right\rangle  = 0$ and $\left\langle {{B_i}(t){B_j}(s)} \right\rangle  = \min (t,s)$. If the covariation matrix of the diffusion components is given by $\Gamma ({{\bm{X}}_t},t) = g({{\bm{X}}_t},t)g{({{\bm{X}}_t},t)^T}$, then, the drift ${f_i}({\bm{X}},t)$ of $i^{th}$-diffusion process and the ${ij}^{th}$-element of the covariation matrix $\Gamma ({\bm{X},t})$ can be estimated as,
\begin{equation}
    \begin{array}{l}
{f_i}({\bm{X}},t) = {\left. {\mathop {\lim }\limits_{\Delta t \to 0} \dfrac{1}{{\Delta t}}E\left[ {{X_i}(t + \Delta t) - {z_i}} \right]} \right|_{{X_k}(t) = {z_k}}}\forall \;k = 1,2, \ldots m\\
{\Gamma _{ij}}({\bm{X}},t) = {\left. {\dfrac{1}{2}\mathop {\lim }\limits_{\Delta t \to 0} \dfrac{1}{{\Delta t}}E\left[ {\left| {{X_i}(t + \Delta t) - {z_i}} \right|\left| {{X_j}(t + \Delta t) - {z_j}} \right|} \right]} \right|_{{X_k}(t) = {z_k}}}\forall \;k = 1,2, \ldots m.
\end{array}
\end{equation}

\section{Conditional probability distributions of the discontinuous Spike and Slab prior (DSS)}\label{appenB}
The DSS prior model is described in the \textit{Methods} section. Drawing sample from the joint distribution is intractable but possible through Markov Chain Monte Carlo methods. In the present work, the Gibbs sampling technique is utilized to draw the sample for the random variables ${\bm{Y}}$, ${\bm{\theta}}$, ${\bm{Z}}$, ${\vartheta _s}$, ${\sigma ^2}$ and ${p_0}$. For drawing samples using Gibbs sampling, the random variables need to be conditioned on other variables. However, due to the DAG structure in Fig. \ref{fig_graph}, the dependencies of the conditional distributions on other random variables can be relaxed to some extent as follows:
\begin{equation}
\begin{aligned}
&p\left(p_{0} \mid \bm{Y}, \bm{\theta}, \bm{Z}, \vartheta_{s}, \sigma^{2}\right)=p\left(p_{0} \mid \bm{Z}\right) \\
&p\left(\vartheta_{s} \mid \bm{Y}, \bm{\theta}, \bm{Z}, p_{0}, \sigma^{2}\right)=p\left(\vartheta_{s} \mid \bm{\theta}, \bm{Z}, \sigma^{2}\right) \\
&p\left(\sigma^{2} \mid \bm{Y}, \bm{\theta}, \bm{Z}, p_{0}, \vartheta_{s}\right)=p\left(\sigma^{2} \mid \bm{Y}, \bm{\theta}, \bm{Z}, \vartheta_{s}\right) \\
&p\left(\bm{\theta} \mid \bm{Y}, \bm{Z}, p_{0}, \vartheta_{s}, \sigma^{2}\right)=p\left(\bm{\theta} \mid \bm{Y}, \bm{Z}, \vartheta_{s}, \sigma^{2}\right) \\
&p\left(\bm{Z} \mid \bm{Y}, \bm{\theta}, p_{0}, \vartheta_{s}, \sigma^{2}\right)=p\left(\bm{Z} \mid \bm{\theta}, p_{0}, \vartheta_{s}, \sigma^{2}\right).
\end{aligned}
\end{equation}
The elements of the weight vector ${\bm{\theta}}$ for which the latent variable $Z_k \ne 1$, becomes an absorbing state in the Markov chain. However, to achieve a stationary distribution, one needs to construct an irreducible Markov chain. Thus, the conditionals over weight vector is eliminated by marginalizing the conditional distributions with respect to the vector ${\bm{\theta}}$. 
\begin{equation}
    \begin{array}{l}
\int {p\left( {{\sigma ^2}\mid {\bm{Y}},{\bm{\theta }},{\bm{Z}},{\vartheta_s}} \right)d{\bm{\theta }}}  = \int {\dfrac{{p\left( {{\sigma ^2},{\bm{Y}},{\bm{\theta }},{\bm{Z}},{\vartheta_s}} \right)}}{{p\left( {{\bm{Y}},{\bm{\theta }},{\bm{Z}},{\vartheta_s}} \right)}}d{\bm{\theta }}} = p\left( {{\sigma ^2}\mid {\bm{Y}},{\bm{Z}},{\vartheta_s}} \right)\\
\int {p\left( {{\bm{Z}}\mid {\bm{Y}},{\bm{\theta }},{p_0},{\vartheta_s},{\sigma ^2}} \right)} d{\bm{\theta }} = \int {\dfrac{{p\left( {{\bm{Z}},{\bm{Y}},{\bm{\theta }},{p_0},{\vartheta_s},{\sigma ^2}} \right)}}{{p\left( {{\bm{Y}},{\bm{\theta }},{p_0},{\vartheta_s},{\sigma ^2}} \right)}}d{\bm{\theta }}} = \dfrac{{p\left( {{\bm{Z}},{\bm{Y}},{p_0},{\vartheta_s},{\sigma ^2}} \right)}}{{p\left( {{\bm{Y}},{p_0},{\vartheta_s},{\sigma ^2}} \right)}} = p\left( {{\bm{Z}}|{\bm{Y}},{p_0},{\vartheta_s},{\sigma ^2}} \right).
\end{array}
\end{equation}

\textbf{The conditional distribution $p\left(p_{0} \mid \bm{Z}\right)$}:
\begin{equation}
    \begin{array}{ll}
        p\left(p_{0} \mid {\bm{Z}}\right) &= p\left(\bm{Z} \mid p_{0}\right) p\left(p_{0}\right) \\
        &\propto\left(\prod_{k=1}^{K} p_{0}^{Z_{k}}\left(1-p_{0}\right)^{1-Z_{k}}\right) \left( p_{0}^{\alpha_{p}-1}\left(1-p_{0}\right)^{\beta_{p}-1} \right) \\
        &\propto p_{0}^{\alpha_{p}+h_{Z}}\left(1-p_{0}\right)^{\beta_{p}+K-h_{Z}},
    \end{array}
\end{equation}
where $h_{Z}=\sum_{k=1}^{K} Z_{k}$. Given the latent vector ${\bm{Z}}$, $p_{0}$ is sampled as, $p_{0} \mid {\bm{Z}} \sim {Beta}\left(\alpha_{p}+h_{z}, \beta_{p}+K-h_{z}\right)$.

\textbf{The conditional distribution $p\left(\vartheta_{s} \mid \bm{\theta}, \bm{Z}, \sigma^{2}\right)$}:
\begin{equation}
    \begin{array}{ll}
        p\left(\vartheta_{s} \mid \bm{\theta}, \bm{Z}, \sigma^{2}\right) & \propto p\left(\bm{\theta} \mid \bm{Z}, \vartheta_{s}, \sigma^{2}\right) p\left(\vartheta_{s}\right) p(\bm{Z}) p\left(\sigma^{2}\right) \\
        & \propto p\left(\bm{\theta} \mid \bm{Z}, \vartheta_{s}, \sigma^{2}\right) p\left(\vartheta_{s}\right)\\
        & \propto \mathcal{N}\left(\bm{\theta}_{r} \mid \mathbf{0}, \sigma^{2} \vartheta_{s} \mathbf{R}_{0, r}\right) \mathcal{I} \mathcal{G}\left(\alpha_{\vartheta}, \beta_{\vartheta}\right) \\
        & \propto \dfrac{1}{\left(\vartheta_{s} \sigma^{2}\right)^{r / 2}\left|\mathbf{R}_{0, r}\right|^{1 / 2}} \exp \left(-\dfrac{\bm{\theta}_{r}^{T} \mathbf{R}_{0, r}^{-1} \bm{\theta}_{r}}{2 \sigma^{2} \vartheta_{s}}\right) \vartheta_{s}^{-\alpha_{\vartheta}-1} \exp \left(-\dfrac{\beta_{\vartheta}}{\vartheta_{s}}\right) \\
        & \propto \vartheta_{s}^{-\left(\alpha_{\vartheta}+\dfrac{r}{2}\right)-1} \exp \left(-\dfrac{\beta_{\vartheta}+\left( {\bm{\theta}_{r}^{T} \mathbf{R}_{0, r}^{-1} \bm{\theta}_{r}}/{2 \sigma^{2}}\right)}{\vartheta_{s}}\right) ,
    \end{array}
\end{equation}
where $r$ is the number of elements of the weight vector ${\bm{\theta}}$ for which the latent variable $Z_k=1$. In the above it is to be noted that $p(\bm{Z})$ and $p\left(\sigma^{2}\right)$ are constants when $\vartheta_{s}$ is sampled. This can be observed from the DAG structure in Fig. \ref{fig_graph}. Thus given $\bm{\theta}$, $\bm{Z}$ and $\sigma^{2}$ and $\vartheta_{s}$ is sampled as, $\vartheta_{s} \mid \bm{\theta}, \bm{Z}, \sigma^{2} \sim \mathcal{I} \mathcal{G}\left(\alpha_{\vartheta}+\dfrac{r}{2}, \beta_{\vartheta}+\dfrac{\bm{\theta}_{r}^{T} \mathbf{R}_{0, r}^{-1} \bm{\theta}_{r}}{2 \sigma^{2}}\right)$.

\textbf{The conditional distribution $p\left(\sigma^{2} \mid \bm{Y}, \bm{\theta}, \bm{Z}, \vartheta_{s}\right)$}:
\begin{equation}
    \begin{aligned}
    p\left(\sigma^{2} \mid \bm{Y}, \bm{Z}, \vartheta_{s}\right) & \propto \int p\left(\bm{Y}, \bm{\theta}, \bm{Z}, \vartheta_{s}, \sigma^{2}\right) d \bm{\theta} \\
    & \propto\left(\int p\left(\bm{Y} \mid \bm{\theta}, \sigma^{2}\right) p\left(\bm{\theta} \mid \bm{Z}, \vartheta_{s}, \sigma^{2}\right) d \bm{\theta}\right) p(\bm{Z}) p\left(\vartheta_{s}\right) p\left(\sigma^{2}\right) \\
    & \propto\left(\int p\left(\bm{Y} \mid \bm{\theta}, \sigma^{2}\right) p\left(\bm{\theta} \mid \bm{Z}, \vartheta_{s}, \sigma^{2}\right) d \bm{\theta}\right) p\left(\sigma^{2}\right) .
    \end{aligned}
\end{equation}
From the DAG structure it can be understood that $p(\bm{Z})$ and $p\left(\vartheta_{s}\right)$ will act as a constant when $\sigma^{2}$ is sampled. Since the $\theta_k$-values corresponding to the spike distribution does not contribute to the basis function selection, the remaining weights belonging to the slab denoted by $\bm{\theta}_{r}$ is used. Then, one can expand the intregrand $p\left({\bm{Y}} \mid \bm{\theta}, \sigma^{2}\right) p\left(\bm{\theta} \mid {\bm{Z}}, \vartheta_{s}, \sigma^{2}\right)$, as,
\begin{equation}
    \begin{aligned}
    &p\left( {{\bm{Y}}\mid {\bm{\theta }},{\sigma ^2}} \right)p\left( {{\bm{\theta }}\mid {\bm{Z}},{\vartheta_s},{\sigma ^2}} \right)\\
     &= \mathcal{N} \left( {{\bm{Y}}|{{\bf{L}}_r}{\bm{\theta} _r},{\sigma ^2}{{\bf{I}}_{N \times N}}} \right)N\left( {{\bm{\theta} _r}|0,{\sigma ^2}{\vartheta_s}{{\bf{R}}_{0,r}}} \right)\\
     &=\dfrac{1}{\left(2 \pi \sigma^{2}\right)^{N / 2}} \exp \left(-\frac{\left({\bm{Y}}-\mathbf{D}_{r} \bm{\theta}_{r}\right)^{T}\left({\bm{Y}}-\mathbf{D}_{r} \bm{\theta}_{r}\right)}{2 \sigma^{2}}\right) \frac{\left(|\mathbf{R}_{0, r}^{-1}|\right)^{1 / 2}}{\left(2 \pi \vartheta_{s} \sigma^{2}\right)^{r / 2}} \exp \left(-\frac{\bm{\theta}_{r}^{T} \mathbf{R}_{0, r}^{-1} \bm{\theta}_{r}}{2 \sigma^{2} \vartheta_{s}}\right)\\
     &= \frac{1}{\left(2 \pi \sigma^{2}\right)^{N / 2}} \frac{\left(|\mathbf{R}_{0, r}^{-1}|\right)^{1 / 2}}{\left(2 \pi \vartheta_{s} \sigma^{2}\right)^{r / 2}} \exp \left(-\frac{\left(\bm{\theta}_{r}-\bm{\mu}\right)^{T} \bm{\Sigma}^{-1}\left(\bm{\theta}_{r}-\bm{\mu}\right)}{2 \sigma^{2}}\right) \exp \left(-\frac{\left(\bm{Y}^{T} \bm{Y}-\bm{\mu}^{T} \bm{\Sigma}^{-1} \bm{\mu}\right)}{2 \sigma^{2}}\right),
    \end{aligned}
\end{equation}
where $\bm{\Sigma}^{-1}=\left(\mathbf{L}_{r}^{T} \mathbf{L}_{r}+\vartheta_{s}^{-1} \mathbf{R}_{0, r}^{-1}\right)$ and $\bm{\mu}=\bm{\Sigma} \mathbf{L}_{r}^{T} {\bm{Y}}.$ Upon integration of the the above expression about ${\bm{\theta}}_{r}$, one obtains:
\begin{equation}
    \begin{aligned}
    p\left(\sigma^{2} \mid \bm{Y}, \bm{Z}, \vartheta_{s}\right) & \propto \frac{1}{\left(\sigma^{2}\right)^{N / 2}} \exp \left(-\frac{\left(\bm{Y}^{T} \bm{Y}-\bm{\mu}^{T} \bm{\Sigma}^{-1} \bm{\mu}\right)}{2 \sigma^{2}}\right) p\left(\sigma^{2}\right) \\
    & \propto \frac{1}{\left(\sigma^{2}\right)^{N / 2}} \exp \left(-\frac{\left(\bm{Y}^{T} \bm{Y}-\bm{\mu}^{T} \bm{\Sigma}^{-1} \bm{\mu}\right)}{2 \sigma^{2}}\right)\left(\sigma^{2}\right)^{-\alpha_{\sigma}-1} \exp \left(-\frac{\beta_{\sigma}}{\sigma^{2}}\right) \\
    & \propto\left(\sigma^{2}\right)^{-\left(\alpha_{\sigma}+{0.5N}\right)-1} \exp \left(-\frac{\beta_{\sigma}+\frac{1}{2}\left(\bm{Y}^{T} \bm{Y}-\bm{\mu}^{T} \bm{\Sigma}^{-1} \bm{\mu}\right)}{\sigma^{2}}\right).
    \end{aligned}
\end{equation}
The random variable $\sigma^{2}$ is sampled as, $\sigma^{2} \mid \bm{Y}, \bm{Z}, \vartheta_{s} \sim \mathcal{I} \mathcal{G}\left(\alpha_{\sigma}+\dfrac{N}{2}, \beta_{\sigma}+\frac{1}{2}\left(\bm{Y}^{T} \bm{Y}-\bm{\mu}^{T} \bf{\Sigma}^{-1} \bm{\mu}\right)\right)$.

\textbf{The conditional distribution $p\left(\bm{\theta} \mid \bm{Y}, \bm{Z}, \vartheta_{s}, \sigma^{2}\right)$}: The elements of the weight vector $\bm{\theta}$ corresponding to the spike distribution i.e. the weights for which the latent variable $z_{k}=0; k=1 \ldots K$, are assigned the value 0. The conditional distribution for the weights belonging to the slab distribution denoted by $\bm{\theta}_{r}$ is derived as follows:
\begin{equation}
    \begin{aligned}
    p\left(\bm{\theta}_{r} \mid \bm{Y}, \vartheta_{s}, \sigma^{2}\right) & \propto \mathcal{N}\left(\bm{Y} \mid \mathbf{L}_{r} \bm{\theta}_{r}, \sigma^{2} \mathbf{I}_{N \times N}\right) \mathcal{N}\left(\bm{\theta}_{r} \mid \mathbf{0}, \sigma^{2} \vartheta_{s} \mathbf{R}_{0, r}\right) \\
    & \propto \exp \left(-\frac{\left(\bm{Y}-\mathbf{L}_{r} \bm{\theta}_{r}\right)^{T}\left(\bm{Y}-\mathbf{L}_{r} \bm{\theta}_{r}\right)}{2 \sigma^{2}}\right) \exp \left(-\frac{\bm{\theta}_{r}^{T} \mathbf{R}_{0, r}^{-1} \bm{\theta}_{r}}{2 \sigma^{2} v_{s}}\right) \\
    & \propto \exp \left(-\frac{\bm{Y}^{T} \bm{Y}+\bm{\theta}_{r}^{T} \bm{\Sigma}^{-1} \bm{\theta}_{r}-2 \bm{\theta}_{r}^{T} \bm{\Sigma}^{-1} \bm{\mu}}{2 \sigma^{2}}\right) \\
    & \propto \exp \left(-\frac{\left(\bm{\theta}_{r}-\bm{\mu}\right)^{T} \bm{\Sigma}^{-1}\left(\bm{\theta}_{r}-\bm{\mu}\right)}{2 \sigma^{2}}\right),
    \end{aligned}
\end{equation}
where $\mathbf{\Sigma}^{-1}=\left(\mathbf{L}_{r}^{T} \mathbf{L}_{r}+\vartheta_{s}^{-1} \mathbf{R}_{0, r}^{-1}\right)$ and $\bm{\mu}=\bm{\Sigma} \mathbf{L}_{r}^{T} \bm{Y}$. Then, the random values of $\bm{\theta}_r$ can be sampled as $\bm{\theta}_{r} \mid \bm{Y}, \vartheta_{s}, \sigma^{2} \sim \mathcal{N}\left(\bm{\mu}, \sigma^{2} \bf{\Sigma}\right)$.

\textbf{The conditional distribution $p\left(\bm{Z} \mid \bm{\theta}, p_{0}, \vartheta_{s}, \sigma^{2}\right)$}: The latent variable $Z_k$ corresponding to the $k^{th}$ element of the weight vector are assigned the value 0 or 1 independently. The conditional probability distributions of $Z_{k}$ are estimated by comparing the probabilities that the $k^{th}$ latent variable $Z_{k}=1$ will assume a value 1 or 0, given the values of $\vartheta_{s}$, $p_{0}$ and remaining values of the latent vector $\bm{Z}_{-k}$. Here, the term $\bm{Z}_{-k}$ denotes the latent vector $\bm{Z}$ whose $k^{th}$ element is removed. Let $u_{k}$ denotes the probability with which the $k^{th}$ latent variable $Z_{k}$ takes a value 1. The probability $u_k$ is then found as,
\begin{equation}
    \begin{aligned}
u_{k} &=\frac{p\left(Z_{k}=1 \mid \bm{Y}, \bm{Z}_{-k}, \vartheta_{s}, p_{0}\right)}{p\left(Z_{k}=1 \mid \bm{Y}, \bm{Z}_{-k}, \vartheta_{s}, p_{0}\right)+p\left(Z_{k}=0 \mid \bm{Y}, \bm{Z}_{-k}, \vartheta_{s}, p_{0}\right)} \\
&=\frac{p\left(\bm{Y} \mid Z_{k}=1, \bm{Z}_{-k}, \vartheta_{s}\right) p\left(Z_{k}=1 \mid p_{0}\right)}{p\left(\bm{Y} \mid Z_{k}=1, \bm{Z}_{-k}, \vartheta_{s}\right) p\left(Z_{k}=1 \mid p_{0}\right)+p\left(\bm{Y} \mid Z_{k}=0, \bm{Z}_{-k}, \vartheta_{s}\right) p\left(Z_{k}=0 \mid p_{0}\right)} \\
&=\frac{p\left(\bm{Y} \mid Z_{k}=1, \bm{Z}_{-k}, \vartheta_{s}\right) p_{0}}{p\left(\bm{Y} \mid Z_{k}=1, \bm{Z}_{-k}, \vartheta_{s}\right) p_{0}+p\left(\bm{Y} \mid Z_{k}=0, \bm{Z}_{-k}, \vartheta_{s}\right)\left(1-p_{0}\right)} \\
&=\frac{p_{0}}{p_{0}+\lambda_{k}\left(1-p_{0}\right)} ,
   \end{aligned}
\end{equation}
where $\lambda_{k}=\dfrac{p\left(\bm{Y} \mid Z_{k}=0, \bm{Z}_{-k}, \vartheta_{s}\right)}{p\left(\bm{Y} \mid Z_{k}=1, \bm{Z}_{-k}, \vartheta_{s}\right)}$. The marginal likelihood function $p\left(\bm{Y} \mid \bm{Z}, \vartheta_{s}\right)$ was derived by integrating out the random variables $\bm \theta$ and $\sigma^{2}$ from the original likelihood function, which follows,
\begin{equation}
p\left(\bm{Y} \mid \bm{Z}, \vartheta_{s}\right) = \int p\left(\bm{Y} \mid \bm{Z}, \vartheta_{s}, \sigma^{2}\right) p\left(\sigma^{2}\right) d \sigma^{2},
\end{equation}
where $p\left(\bm{Y} \mid \bm{Z}, \vartheta_{s}, \sigma^{2}\right)$ is obtained by marginalizing the likelihood function with respect to the variable $\bm{\theta}$. Considering only the weights whose corresponding latent variables $Z_k=0$, denoted by $\bm{\theta_r}$, the probability $p\left(\bm{Y} \mid \bm{Z}, \vartheta_{s}, \sigma^{2}\right)$ is obtained as,
\begin{equation}
    \begin{aligned}
p\left(\bm{Y} \mid \bm{Z}, \vartheta_{s}, \sigma^{2}\right) 
=& \int p\left(\bm{Y}, \bm{\theta} \mid \bm{Z}, \vartheta_{s}, \sigma^{2}\right) d \theta \\
=& \int p\left(\bm{Y} \mid \bm{\theta}_{r}, \sigma^{2}\right) p\left(\bm{\theta}_{r} \mid \vartheta_{s}, \sigma^{2}\right) d \bm{\theta}_{r} \\
=& \int \mathcal{N}\left(\bm{Y} \mid \mathbf{L}_{r} \bm{\theta}_{r}, \sigma^{2} \mathbf{I}_{N \times N}\right) \mathcal{N}\left(\bm{\theta}_{r} \mid 0, \sigma^{2} \vartheta_{s} \mathbf{R}_{0, r}\right) d \bm{\theta}_{r} \\
=& \int \frac{1}{\left(2 \pi \sigma^{2}\right)^{N / 2}} \exp \left(-\frac{\left(\bm{Y}-\mathbf{L}_{r} \bm{\theta}_{r}\right)^{T}\left(\bm{Y}-\mathbf{L}_{r} \bm{\theta}_{r}\right)}{2 \sigma^{2}}\right) \frac{\left(|\mathbf{R}_{0, r}^{-1}|\right)^{1 / 2}}{\left(2 \pi \vartheta_{s} \sigma^{2}\right)^{r / 2}} \exp \left(-\frac{\bm{\theta}_{r}^{T} \mathbf{R}_{0, r}^{-1} \bm{\theta}_{r}}{2 \sigma^{2} \vartheta_{s}}\right) d \bm{\theta}_{r} \\
=& \frac{1}{\left(2 \pi \sigma^{2}\right)^{N / 2}} \frac{\left(|\mathbf{R}_{0, r}^{-1}|\right)^{1 / 2}\left(|\mathbf{\Sigma}^{-1}|\right)^{1 / 2}}{\left(\vartheta_{s}\right)^{r / 2}} \exp \left(-\frac{\left(\bm{Y}^{T} \bm{Y}-\bm{\mu}^{T} \mathbf{\Sigma} \bm{\mu}\right)}{2 \Sigma^{2}}\right),
\end{aligned}
\end{equation}
where $\mathbf{\Sigma}=\left(\mathbf{L}_{r}^{T} \mathbf{L}_{r}+\vartheta_{s}^{-1} \mathbf{R}_{0, r}^{-1}\right)^{-1}$ and $\bm{\mu}=\mathbf{\Sigma} \mathbf{L}_{r}^{T} \bm{Y}$. The operator $|\cdot|$ denotes the determinant. With the above result, $p\left(\bm{Y} \mid \bm{Z}, \vartheta_{s}\right)$ can be derived as,
\begin{equation}
    \begin{aligned}
p\left(\bm{Y} \mid \bm{Z}, \vartheta_{s}\right)
=& \frac{\left(|\mathbf{R}_{0, r}^{-1}|\right)^{1 / 2}\left(|\mathbf{\Sigma}^{-1}|\right)^{1 / 2}}{(2 \pi)^{N / 2}\left(\vartheta_{s}\right)^{r / 2}} \int \frac{1}{\left(\sigma^{2}\right)^{N / 2}} \exp \left(-\frac{\left(\bm{Y}^{T} \bm{Y}-\bm{\mu}^{T} \mathbf{\Sigma} \bm{\mu}\right)}{2 \sigma^{2}}\right) \mathcal{I} \mathcal{G}\left(\alpha_{\sigma}, \beta_{\sigma}\right) d \sigma^{2} \\
=& \frac{\left(|\mathbf{R}_{0, r}^{-1}|\right)^{1 / 2}\left(|\mathbf{\Sigma}^{-1}|\right)^{1 / 2}}{(2 \pi)^{N / 2}\left(\vartheta_{s}\right)^{r / 2}} \frac{\left(\beta_{\sigma}\right)^{\alpha_{\sigma}}}{\Gamma\left(\alpha_{\sigma}\right)} \int \frac{1}{\left(\sigma^{2}\right)^{\alpha_{\sigma}+N / 2+1} \exp \left(-\frac{\beta_{\sigma}+\frac{1}{2}\left(\bm{Y}^{T} \bm{Y}-\bm{\mu}^{T} \mathbf{\Sigma} \bm{\mu}\right)}{\sigma^{2}}\right)} d \sigma^{2} \\
=& \frac{\left(|\mathbf{R}_{0, r}^{-1}|\right)^{1 / 2}\left(|\mathbf{\Sigma}^{-1}|\right)^{1 / 2}}{(2 \pi)^{N / 2}\left(\vartheta_{s}\right)^{r / 2}} \frac{\left(b_{\sigma}\right)^{\alpha_{\sigma}}}{\Gamma\left(\alpha_{\sigma}\right)} \frac{\Gamma\left(\alpha_{\sigma}+\frac{N}{2}\right)}{\left(\beta_{\sigma}+\frac{1}{2}\left(\bm{Y}^{T} \bm{Y}-\bm{\mu}^{T} \mathbf{\Sigma} \bm{\mu}\right)\right)^{\left(\alpha_{\sigma}+\frac{N}{2}\right)}},
\end{aligned}
\end{equation}
where the operator $\Gamma(\cdot)$ denotes the Gamma function. To summarise, the random variables $Z_{k}$ can be computed from the Bernoulli distribution with the parameter $u_{k}$ as, $Z_{k}; k= 1, \ldots, K \quad \mid \bm{Y}, \vartheta_{s}, p_{0} \sim {Bern}\left(u_{k}\right)$, where the marginalised likelihood function is obtained as,
\begin{equation}
p\left(\bm{Y} \mid \bm{Z}, \vartheta_{s}\right)= \begin{cases}\dfrac{\Gamma\left(\alpha_{\sigma}+\dfrac{N}{2}\right)}{(2 \pi)^{N / 2} \dfrac{\left(\beta_{\sigma}\right)^{\alpha_{\sigma}}}{\Gamma\left(\alpha_{\sigma}\right)}} \dfrac{1}{\left(\beta_{\sigma}+\dfrac{1}{2}\left(\bm{Y}^{T} \bm{y}\right)\right)^{\left(\alpha_{\sigma}+\dfrac{N}{2}\right)}} \quad \text{, when all $\{Z_{k}; k=1, \ldots, K \}$ = 0} \\ \dfrac{\Gamma\left(a_{\sigma}+\dfrac{N}{2}\right)}{(2 \pi)^{N / 2}\left(\vartheta_{s}\right)^{r / 2}} \dfrac{\left(\beta_{\sigma}\right)^{\alpha_{\sigma}}}{\Gamma\left(\alpha_{\sigma}\right)} \dfrac{\left(|\mathbf{R}_{0, r}^{-1}|\right)^{1 / 2}\left(|\mathbf{\Sigma}^{-1}|\right)^{1 / 2}}{\left(\beta_{\sigma}+\dfrac{1}{2}\left(\bm{Y}^{T} \bm{Y}-\bm{\mu}^{T} \mathbf{\Sigma} \bm{\mu}\right)\right)^{\left(\alpha_{\sigma}+\dfrac{N}{2}\right)}}\quad \text{, otherwise.} \end{cases}
\end{equation}

\section*{Acknowledgements} T. Tripura acknowledges the financial support received from the Ministry of Human Resource Development (MHRD), India in form of the Prime Minister's Research Fellows (PMRF) scholarship. S. Chakraborty acknowledges the financial support received from Science and Engineering Research Board (SERB) through grant no. SRG/2021/000467 and seed grant received from IIT Delhi.

\section*{Declarations}

\subsection*{Funding} The corresponding author received funding from IIT Delhi in form of seed grant.

\subsection*{Conflicts of interest} The authors declare that they have no conflict of interest.

\subsection*{Availability of data and material} The datasets generated during and/or analysed during the current study are available from the corresponding author on reasonable request.

\subsection*{Code availability} The Python codes written for this work are available from the corresponding author on reasonable request.

% % Bibliography
% \bibliographystyle{unsrt}  
% \bibliography{mybibfile}  

\end{document}